\documentclass[10pt,twocolumn,final]{IEEEtran}

\usepackage{rotating}
\usepackage{cite}
\usepackage[cmex10]{amsmath}
\usepackage{amsthm}
\usepackage{amssymb}
\usepackage{balance}
\usepackage{color}
\usepackage{graphicx}
\usepackage[font=footnotesize]{caption}
\usepackage[font=footnotesize]{subcaption}

\newtheorem{corollary}{Corollary}

\newcommand{\beqa}{\begin{eqnarray}}
\newcommand{\eeqa}{\end{eqnarray}}
\newcommand{\beqas}{\begin{eqnarray*}}
	\newcommand{\eeqas}{\end{eqnarray*}}

\DeclareMathOperator*{\maximize}{maximize}
\DeclareMathOperator*{\minimize}{minimize}
\DeclareMathOperator*{\st}{subject\;to}
\usepackage[ruled,vlined,linesnumbered]{algorithm2e}

\newtheorem{theorem}{Theorem}
\newtheorem{proposition}{Proposition}
\newtheorem{property}{Property}
\newcommand{\bs}[1]{\boldsymbol{#1}}
\newcommand{\mc}[1]{\mathcal{#1}}

\newcommand{\mb}[1]{\mathbf{#1}}
\newcommand{\mr}[1]{\mathrm{#1}}
\newcommand{\tr}{\mathrm{Tr}}

\title{Optimal Dynamic Point Selection for Power Minimization in Multiuser Downlink CoMP
\thanks{Duy H. N. Nguyen is with the Department
of Electrical and Computer Engineering, San Diego State University, San Diego, CA 92182, USA (e-mail: duy.nguyen@sdsu.edu).} 
\thanks{Long B. Le is with INRS-EMT, Universit\'e du Qu\'ebec, Montr\'eal, QC, Canada, H5A 1K6 (email: long.le@emt.inrs.ca).}
\thanks{Tho Le-Ngoc is with the Department
of Electrical and Computer Engineering, McGill University,
3480 University Street, Montreal, QC, Canada, H3A 0E9 (email: tho.le-ngoc@mcgill.ca).}}
\author{Duy H. N. Nguyen, \emph{Member, IEEE}, Long B. Le, \emph{Senior Member, IEEE}, and \\ Tho Le-Ngoc, \emph{Fellow, IEEE}}
\begin{document}
\maketitle
\begin{abstract}
This paper examines a CoMP system where multiple base-stations (BS) employ coordinated beamforming to serve multiple mobile-stations (MS). Under the dynamic point selection mode, each MS can be assigned to only one BS at any time. This work then presents a solution framework to optimize the BS associations and coordinated beamformers for all MSs. With target signal-to-interference-plus-noise ratios at the MSs, the design objective is to minimize either the weighted sum transmit power or the per-BS transmit power margin. Since the original optimization problems contain binary variables indicating the BS associations, finding their optimal solutions is a challenging task. To circumvent this difficulty, we first relax the original problems into new optimization problems by expanding their constraint sets. Based on the nonconvex quadratic constrained quadratic programming framework, we show that these relaxed problems can be solved optimally. Interestingly, with the first design objective, the obtained solution from the relaxed problem is also optimal to the original problem. With the second design objective, a suboptimal solution to the original problem is then proposed, based on the obtained solution from the relaxed problem. Simulation results show that the resulting jointly optimal BS association and beamforming design significantly outperforms fixed BS association schemes.
\end{abstract}


\section{Introduction}
\begin{figure}[!thb]
	\centering
	\begin{subfigure}[b]{0.42\textwidth}
		\includegraphics[width=\textwidth]{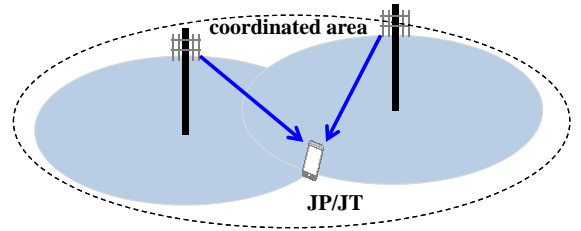}
		\caption{Joint Processing/Joint Transmission (JP/JT): a MS is served
			simultaneously by two BSs.}
		\label{fig:JP}
	\end{subfigure}
	~~
	\begin{subfigure}[b]{0.42\textwidth}
		\includegraphics[width=\textwidth]{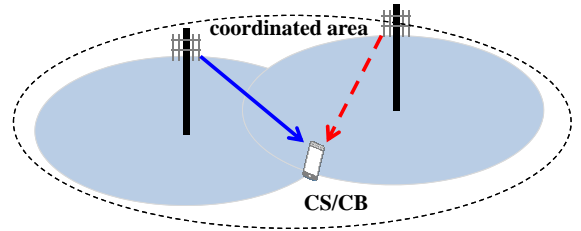}
		\caption{Coordinated Scheduling/Coordinated Beamforming (CS/CB): a MS's signal (solid blue arrow)
			is transmitted by one BS and the interference (red dashed arrow) coming
			from the other BS is coordinated.}
		\label{fig:CB}
	\end{subfigure}
	~~
	\begin{subfigure}[b]{0.42\textwidth}
		\includegraphics[width=\textwidth]{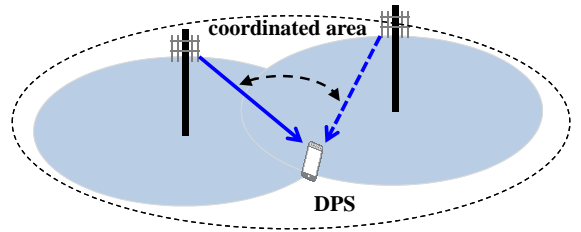}
		\caption{Dynamic Point Selection (DPS): a MS is served by one single BS at
			any time and MS-BS association can be changed accordingly to the channel conditions.}
		\label{fig:DPS}
	\end{subfigure}
	\caption{Example operations of different CoMP modes.}
	\label{fig:CoMP-mode}
\end{figure}

To improve the spectral efficiency, current designs of wireless networks adopt universal
frequency reuse where all the cells can share the same radio spectrum resources.
However, universal frequency reuse comes at the cost of severe intercell interference (ICI),
especially at cell-edge mobile stations (MS). In the 3GPP LTE-Advanced standard,
coordinated multi-point transmission/reception (CoMP) is considered as an enabling technique
to actively deal with the ICI \cite{3GPP}. In CoMP, the coverage, throughput and efficiency of the multicell system can be significantly
improved by fully coordinating and optimizing the concurrent transmissions from multiple base-stations (BS)
to the MSs \cite{3GPP,Duy-Springer14}. Depending on the level of coordination among the coordinated cells,
a CoMP system can operate under different modes, namely joint processing/joint transmission (JP/JT),
dynamic point selection (DPS), and coordinated scheduling/coordinated beamforming (CS/CB)  \cite{3GPP,Maattanen2012},
as illustrated in Fig. \ref{fig:CoMP-mode}.

In the JP/JT mode (Fig. \ref{fig:JP}), the antennas of a cluster of coordinated BSs
form a large single antenna array \cite{Valenzuela06,Gesbert10}.
The signals intended for a particular MS are simultaneously transmitted from multiple BSs across cell sites.
Thus, JP/JT offers the benefit of large-scale BS cooperation \cite{Zakhour-Hanly-TIT2012,Sanguinetti-Debbah-TWC2016}.
Asymptotic performance of JP/JT has been analyzed in recent works through the large system analysis of coordinated
multicell systems \cite{Zakhour-Hanly-TIT2012,Sanguinetti-Debbah-TWC2016,Zhang-Wen-Gao-Wong-TWC2013}.
Although JP/JT can exploit the best performance from the CoMP system, it is the most complex mode 
in terms of signaling and synchronization among the BSs \cite{3GPP}. Per the 3GPP LTE-Advanced Release 11, the JP/JT mode
is normally assumed to be ``coherent'', meaning that co-phasing of the signs from different coordinated
transmission points is performed by means of precoding at the transmitter \cite{3GPP}. 
Thus, implementing JP/JT will need a high-resolution adjustable analog delay to each coordinated BS
to cope with the delay variations. For this reason, it is difficult to fully realize the 
potential performance gains of coherent JP/JT, which may limit their applicability only to BSs connected by a fast backhaul 
\cite{Lee-Clerckx-CMag2012,Lee12b}. In addition, coherent JP/JT requires inter-point phase information
as part of the channel state information (CSI) feedback from multiple points \cite{3GPPTR36.819}.

The CS/CB mode accounts for the least complex CoMP mode. In CS/CB (Fig. \ref{fig:CB}), 
the signal to a single MS is transmitted from the serving cell only \cite{3GPP}. 
However, the beamforming functionality is dynamically coordinated between the BSs to control/reduce the ICI \cite{3GPP,Maattanen2012,WeiYu10}. Optimal beamforming design for CoMP system under CS/CB mode can be
obtained from joint optimization \cite{WeiYu10,Zakhour-Hanly-TIT2012,Sanguinetti-Debbah-TWC2016,Huang-Ottersten-TSP2011,
He-Huang-TWC2012,Huang-Ottersten-TWC2012,Huang-Tan-Rao-TWC2013} or game theory \cite{Duy-TSP11}.
To effectively coordinate the inter-cell interference, CS/CB requires CSI feedback from multiple points \cite{3GPPTR36.819}.
However, by exploiting channel reciprocality \cite{3GPPTR36.819}, optimal downlink CS/CB
can be implemented if a BS knows the CSI only to its connected MSs \cite{WeiYu10}.

In DPS mode (Fig. \ref{fig:DPS}), the MS, at any one time, is being associated to
a single BS. However, this single associated BS can dynamically change from time-frame
to time-frame within a set of possible BSs inside the cluster \cite{3GPP,Maattanen2012,Agrawal-WCNC2014}. 
CoMP DPS provides a good trade-off between the transmission algorithm complexity,
system performance and backhaul overhead, in comparison to JP/JT and CS/CB \cite{Feng-VTC2010}.
In fact, the synchronization issue and the requirement of fast backhaul communications 
can be alleviated in the DPS mode, compared to the JP/JT mode.
In DPS, each MS's data has to be available at all the possible
BSs ready for selection. In addition, the beamforming functionality is still needed to
coordinate the transmission across the BSs for interference control \cite{3GPP}. 
To facilitate the interference control, DPS demands similar CSI feedback as CS/CB such that no inter-point phase
information is required \cite{3GPPTR36.819}. 
In fact, when the user-BS association is determined, the DPS mode becomes the CS/CB mode. 
Compared to CS/CB, DPS offers the advantage of site selection
diversity, since DPS can provide a ``soft-handoff'' solution to among the coordinated BSs to quickly
switch the best BS for association for each MS.
However, it is not clear how a joint BS association strategy and beamforming design in the DPS mode can be
optimally determined to maximize the performance of the CoMP system.
In this paper, we are interested in jointly optimizing the BS association strategy
and linear beamforming design for a CoMP downlink system under the DPS mode.
With a set of target signal-to-interference-plus-noise ratios (SINR) at the MSs, 
our design objective is to minimize either: i.) the weighted sum transmit power across the BSs 
or ii.) the per-BS transmit power margin.

\subsection{Related Works}
Designing multicell beamformers under the CS/CB mode has attracted
a lot of research attention, such as \cite{WeiYu10,Huang-Tan-Rao-TWC2013,Zakhour-Hanly-TIT2012,
Sanguinetti-Debbah-TWC2016,Huang-Ottersten-TSP2011,He-Huang-TWC2012,Huang-Ottersten-TWC2012,Duy-TSP11}. Uplink-downlink duality and iterative fixed-point iteration have been successfully exploited in \cite{WeiYu10,Huang-Ottersten-TSP2011,He-Huang-TWC2012,Huang-Ottersten-TWC2012,Huang-Tan-Rao-TWC2013} to obtain optimal beamformers to either minimize the 
sum transmit power at the BSs or maximize the minimum SINR at the MSs. Different to 
these previous studies, part of this work examines the application of uplink-downlink duality to optimize the multicell beamformers under the DPS mode.

While the problem of joint BS association and beamforming design/power control in uplink transmission has been intensively studied \cite{Hanly95,Farrokhi98,Yates95b},
the counterpart problem in downlink transmission is not well understood.
There are few prior studies in literature which deal with this downlink transmission problem.
In \cite{Farrokhi97}, the problem for downlink transmission has been investigated
for the case of power control (not including beamforming design). 
It is stated in \cite{Farrokhi97} that there is no Pareto-optimal solution for the problem of joint 
BS association and beamforming design in the downlink. The work in \cite{Stridh06} 
tackled the problem of joint downlink beamforming, power control, and access point allocation
in a congested system. In \cite{Duy-IET13}, the joint optimization of BS association and beamforming design was examined 
and a relaxing-and-rounding technique was proposed as a suboptimal solution to the binary variables indicating
the BS association. Recent works in \cite{Fallgren11,MingyiHong-JSAC13,Sanjabi14} proposed joint BS association 
and power allocation/beamforming design strategies to maximize the multicell system throughput.
In another work \cite{Sun-Hong-Luo-2012}, the problem of joint BS assignment and power allocation 
for maximizing the minimum rate in a single-input single-output (SISO) interference channel was investigated.
A two-stage algorithm was proposed to iteratively find the BS assignment and power allocation for the users \cite{Sun-Hong-Luo-2012}.
In contrast to these works, our formulation and solution framework are to attain Pareto-optimal joint
BS association and beamforming design strategies with guaranteed SINRs at the MSs.

In the context of finding the optimal beamforming design for power minimization, 
the optimization can be formulated as a nonconvex quadratic constrained quadratic programming (QCQP) problem \cite{Bengt01}.
The nonconvex QCQP is then solved indirectly via convex semi-definite programming (SDP) relaxation \cite{Bengt01} or
a transformation into a convex second-order conic programming (SOCP) problem \cite{Wiesel06,WeiYu07,WeiYu10}. 
It will be shown later in this paper that it is not possible to transform the problems under consideration
into a SOCP. Thus, we rely on recent developments in nonconvex QCQP \cite{HoangTuy13} in joinly devising
the optimal BS association strategy and beamforming design.



\subsection{Contributions of This Paper}
In this work, we formulate the joint BS association and beamforming design problems
as mixed integer programs, which contain the binary variables indicating the BS associations.
To circumvent the difficulty in dealing with the binary variables and
devise optimal joint BS association and beamforming designs, our proposed solution approaches,
which also account for the main contributions of this paper, are as follows:
\begin{itemize}
	\item We propose a relaxation method to solve these original mixed integer programs by 
	relaxing all the binary variables to $1$ and focusing on optimizing the
	beamformers. These relaxed optimization problems are shown to be nonconvex QCQP.
	Our analysis based the QCQP solution framework then shows that the relaxed problems can be solved optimally. 
	\item Under the design objective of minimizing the weighted sum transmit power, 
	the obtained solution from the relaxed problem is also optimal to the original problem.
	Specifically, this solution indicates both the optimal BS association strategy and the optimal beamforming design
	for all MSs. 	Our proposed framework also indicates that 
	any Pareto-optimal solution can be obtained by properly adjusting the weight factors in the objective function of sum transmit power.
	\item We propose two solution approaches based on the Lagrangian duality and the dual uplink problem
	to find the optimal solution. Via the dual uplink problem, we propose a distributed algorithm 
	to obtain the optimal joint BS association and beamforming design. We show that
	the DPS can be optimally implemented when a BS knows the CSI only to users within its serving user set.
	\item Under the design objective of minimizing the per-BS transmit power margin, 
	the optimality of the relaxed problem's solution to the original problem is not always observed.
	Nevertheless, based on the obtained solution from the relaxed problem,
	a suboptimal solution to the original problem is then proposed. We observe
	that the performance gap between the suboptimal solution to the optimal one is negligible in simulations.
	Simulation results also show that the resulting optimal joint BS association and
	beamforming design can significantly improve the performance of the CoMP system.
\end{itemize}

\emph{Notations}: Superscripts $(\cdot)^T$, $(\cdot)^*$, $(\cdot)^{H}$ stand for
transpose, complex conjugate, and complex conjugate transpose operations, respectively;
upper-case bold face letters are used to denote matrices whereas lower-case bold face letters are used to denote column vectors;
$\mathrm{diag}(d_1,d_2,\ldots,d_M)$ denotes an $M\times M$ diagonal matrix with diagonal elements $d_1,d_2,\ldots,d_M$;
$[\cdot]_{i,j}$ denotes the $(i,j)$ element of the matrix argument;
$x^{\star}$ indicates the optimal value of the variable $x$;
$\mb{A} \succeq \mb{B}$ (and $\mb{A} \succ \mb{B}$) is to indicate the matrix inequality (and strict matrix inequality)
defined on the cone of nonnegative definite matrices;
$\mb{A} \nsucc \mb{B}$ is to denote that $\mb{A} - \mb{B}$ is a semi-definite and singular matrix;
$|x|$ denotes the absolute value of the scalar number $x$ whereas $|\mc{X}|$ denotes the cardinality 
of the set $\mc{X}$; 
$\mathbb{C}$ and $\mathbb{R}$ denote the sets of complex and real numbers, respectively.

\section{System Model and Problem Formulation} \label{sect-2}
\subsection{System Model}
We consider the multiuser downlink transmission in a multicell network
consisting of $Q$ BSs and $K$ MSs operating on a same frequency band.
Denote $\mc{Q}$ and $\mc{K}$ as the set of BSs and MSs, respectively.
It is assumed that each BS is equipped with $M$ transmit antennas and each MS is equipped
with a single receive antenna. In each cell, the BS multiplexes and concurrently sends multiple data streams to
multiple MSs. However, each MS can be only associated with one BS at any time.
In the downlink transmission to a particular MS, say MS-$i$, its received signal $y_i$
can be modeled as
\begin{equation}
y_i = \sum_{q=1}^Q \mb{h}_{iq}^H \mb{x}_{q} + z_{i},
\end{equation}
where $\mb{x}_{q} \in \mathbb{C}^{M\times 1}$ is the transmitted signal
at BS-$q$, $\mb{h}^*_{iq}\in \mathbb{C}^{M\times 1}$ represents the channel from BS-$q$
to MS-$i$, and $z_i$ is the AWGN with a power spectral density $\sigma^2$.

Let $\mc{Q}_i \subset \mc{Q}$ be the cluster of coordinated BSs serving user-$i$,
and let $\mc{K}_q$ be the serving user set by BS-$q$.
Let us define binary variables $a_{iq}, i\in\mc{K},q \in \mc{Q}_i$ to represent the association between
MS-$i$ and BS-$q$. More specifically, the binary variable $a_{iq} = 1$ if and only if BS-$q \in \mc{Q}_i$
is assigned to serve MS-$i$.
By means of linear beamforming, the transmitted signal at BS-$q$ can be formulated as
\begin{equation}
\mb{x}_q = \sum_{i\in\mc{K}_q} \mb{w}_{iq}u_i,
\end{equation}
where $\mb{w}_{iq} \in \mathbb{C}^{M\times 1}$ is the beamforming vector and $u_{i}$ is a complex scalar
representing the signal intended for MS-$i$. Without loss of generality, let $\mathbb{E}[|u_{i}|] = 1$.
Clearly, if $a_{iq} = 0$, $\mb{w}_{iq}$ needs to be set at all-$0$ vector.
If BS-$q$ with $q\in\mc{Q}_i$ is selected to serve MS-$i$, the SINR at MS-$i$ is then given by
\begin{eqnarray} \label{SINR-1}
\mathrm{SINR}_{iq} = \frac{\big|\mb{h}_{iq}^H\mb{w}_{iq}\big|^2}{
\sum\limits_{j\neq i}^K\sum\limits_{r\in\mc{Q}_j}\left|\mb{h}_{ir}^H\mb{w}_{jr}\right|^2 + \sigma^2}.
\end{eqnarray}

\subsection{Problem Formulation}
We first consider the joint BS association and beamforming design with the design objective of minimizing the weighted sum transmit power across the
BSs with a set of target SINRs at the MSs. Let $w_q$ be the positive weight for the transmit
power at BS-$q$. The optimization problem is then stated as
\begin{eqnarray}
\mc{P}_1\,:\quad&\underset{\{a_{iq}\},\{\mb{w}_{iq}\}}{\minimize}&\; \sum_{q=1}^Q w_q  \sum_{i\in\mc{K}_q} \left\|\mb{w}_{iq}\right\|^2 \\
&\st&\; \sum_{q \in \mc{Q}_i} a_{iq}\mr{SINR}_{iq} \geq \gamma_i,\forall i \nonumber \\
&&\; a_{iq} = \{0,1\},\forall q\in\mc{Q}_i,\forall i \nonumber \\
&&\; \sum_{q\in\mc{Q}_i} a_{iq} = 1, \forall i, \nonumber
\end{eqnarray}
where the last constraint is to ensure that only one BS in $\mc{Q}_i$ will be associated with MS-$i$.

\emph{Remark 1:} With a predetermined BS association strategy (known $a_{iq}$'s), problem $\mc{P}_1$ becomes the CS/CB design problem, whose optimal solution is readily obtainable \cite{WeiYu10}.
In this case, the SINR constraints can be cast as convex second-order conic (SOC) constraints,
which effectively transforms the optimization problem into a convex one.
However, with the dynamic BS association strategy, the presence of binary variables $a_{iq}$,
problem $\mc{P}_1$ is a nonconvex mixed integer program, which is NP-hard \cite{Aardal02}.
In fact, an exhaustive search for the optimal BS association has exponential complexity and is
impractical for implementation.

One common method to solve a mixed integer program is relaxing the discrete variables into continuous ones
\cite{Joshi09}. 
In this work, we take a completely different approach by setting the binary variables $a_{iq}$'s to 1s.
More precisely, we consider the following optimization problem:
\vspace{-0.1cm}
\begin{eqnarray}
\mc{P}'_1\,:\quad&\underset{\{\mb{w}_{iq}\}}{\minimize}&\; \sum_{q=1}^Q w_q  \sum_{i\in\mc{K}_q}\left\|\mb{w}_{iq}\right\|^2 \\
&\st&\; \sum_{q \in \mc{Q}_i} \frac{\big|\mb{h}_{iq}^H\mb{w}_{iq}\big|^2}{
\sum\limits_{j\neq i}^K\sum\limits_{r\in\mc{Q}_j}\left|\mb{h}_{ir}^H\mb{w}_{jr}\right|^2 + \sigma^2} \geq \gamma_i,\forall i \nonumber.
\end{eqnarray}

\begin{theorem} \label{prop-1}
The minimum weighted sum transmit power obtained from solving problem $\mc{P}'_1$ 
is a lower-bound to that obtained from solving problem $\mc{P}_1$.
\end{theorem}

\begin{IEEEproof}
Suppose that $(\{a_{iq}^{\star}\},\{\mb{w}_{iq}^{\star}\})$ is the optimal solution to the original joint BS
and beamforming problem $\mc{P}_1$. 
From the solution $(a_{iq}^{\star},\mb{w}_{iq}^{\star})$, we denote $q_i \in \mc{Q}_i$ as the BS associated with MS-$i$. 
Since a MS can only be assigned to one BS, we have $a_{iq_i}^{\star} = 1$, $a_{iq}^{\star} = 0,\forall q\neq q_i$,
and $\mb{w}_{iq_i}^{\star} \neq \mb{0}$ and $\mb{w}_{iq}^{\star} = \mb{0},\forall q\in\mc{Q}_i,q\neq q_i$.
In addition, $\{\mb{w}_{iq_i}^{\star}\}$ must be an optimal solution to the following problem
\vspace{-0.1cm}
\begin{eqnarray}
\bar{\mc{P}}_1\,:\quad &\underset{\{\mb{w}_{iq_i}\}}{\minimize}&\; \sum_{i=1}^K w_{q_i}\left\|\mb{w}_{iq_i}\right\|^2 \\
&\st&\; \frac{\big|\mb{h}_{iq_i}^H\mb{w}_{iq_i}\big|^2}{
	\sum\limits_{j\neq i}^K\big|\mb{h}_{iq_j}^H\mb{w}_{jq_j}\big|^2 + \sigma^2} \geq \gamma_i,\forall i,\nonumber
\end{eqnarray}
where $\big|\mb{h}_{iq_j}^H\mb{w}_{jq_j}\big|^2$ is the interference induced by the BS connected to
MS-$j$, \emph{i.e.}, $q_j$, to MS-$i$.
If additional constraints $\mb{w}_{iq} = \mb{0},\forall q\in\mc{Q}_i,q\neq q_i,\forall i$ are introduced to problem $\bar{\mc{P}}_1$,
we will have the following optimization problem
\vspace{-0.1cm}	
\begin{eqnarray}
\hat{\mc{P}}_1\,:\quad &\underset{\{\mb{w}_{iq}\}}{\minimize}&\; \sum_{q=1}^Q w_q  \sum_{i\in\mc{K}_q}\left\|\mb{w}_{iq}\right\|^2 \\
&\st&\;  \frac{\sum_{q \in \mc{Q}_i}\big|\mb{h}_{iq}^H\mb{w}_{iq}\big|^2}{
	\sum\limits_{j\neq i}^K\sum\limits_{r\in\mc{Q}_j}\left|\mb{h}_{ir}^H\mb{w}_{jr}\right|^2 + \sigma^2} \geq \gamma_i,\forall i \nonumber \\
&&\; \mb{w}_{iq} = \mb{0}, \forall q\in\mc{Q}_i, q\neq q_i, \forall i. \nonumber
\end{eqnarray}
Due to the additional constraints $\mb{w}_{iq} = \mb{0},\forall q\in\mc{Q}_i,q\neq q_i,\forall i$, 
the objective functions of problems $\bar{\mc{P}}_1$ and  $\hat{\mc{P}}_1$ are the same. 
In addition, the numerators in the SINR constraints in the two problems are the same, so
are the denominators. Thus, problem $\hat{\mc{P}}_{1}$ must yield \emph{the same solution} as problem $\bar{\mc{P}}_1$, \emph{i.e.}, same optimal point.
If the additional constraints $\mb{w}_{iq} = \mb{0},\forall q\in\mc{Q}_i,q\neq q_i,\forall i$ are
now removed from problem $\hat{\mc{P}}_1$, we will have problem $\mc{P}_1'$.
Since problem $\mc{P}'_1$ has a larger feasibility region than problem $\hat{\mc{P}}_1$, the minimum
point of problem $\mc{P}'_1$ must not exceed that of problem $\hat{\mc{P}}_1$. As a result,
the optimal point of problem $\mc{P}'_1$ is a \emph{lower bound} to the optimal point of the original
problem $\mc{P}_1$.	
%
\end{IEEEproof}

\begin{corollary} \label{coro-1}
If problem $\mc{P}_1$ is feasible, problem $\mc{P}'_1$ is also feasible.
Conversely, the infeasibility of problem $\mc{P}'_1$ also indicates the infeasibility of problem $\mc{P}_1$.
\end{corollary}

\begin{corollary} \label{coro-2}
	Given that $\{\mb{w}_{iq}^{\star}\}$ with $i\in\mc{K}, q\in\mc{Q}_i$ is the set of optimal beamformers
	obtained from solving problem $\mc{P}'_1$, 
	if there exists $q_i \in\mc{Q}_i$
	such that $\mb{w}^{\star}_{iq_i} \neq \mb{0}$ and $\mb{w}^{\star}_{iq} = \mb{0},\forall q\neq q_i$,
	then $\{\mb{w}_{iq}^{\star}\}$ is also the optimal solution
	to problem $\mc{P}_1$.
\end{corollary}
\begin{IEEEproof}
This corollary comes directly from Theorem \ref{prop-1} and its proof.
The optimal BS association for MS-$i$ is then given by the BS index $q_i\in\mc{Q}_i$
corresponding to $\mb{w}^{\star}_{iq_i} \neq \mb{0}$. Moreover,
$\mb{w}^{\star}_{iq_i}$ is also the optimal beamforming vector for MS-$i$.
\end{IEEEproof}


In the following sections, we focus on solving problem $\mc{P}'_1$. 
It is noted the
SINR constraint in problem $\mc{P}_1'$ can be restated as
\vspace{-0.2cm}
\begin{eqnarray} \label{quadratic}
\sum_{q \in \mc{Q}_i} \big|\mb{h}_{iq}^H\mb{w}_{iq}\big|^2 \geq \gamma_{i}
\sum\limits_{j\neq i}^K\sum\limits_{r\in\mc{Q}_j} \big|\mb{h}_{ir}^H\mb{w}_{jr}\big|^2 + \gamma_i\sigma^2.
\end{eqnarray}
If there is only \emph{one} term on the left hand side of the above inequality constraint, 
say $\big|\mb{h}_{i{q}_i}^H\mb{w}_{i{q}_i}\big|^2$, one can assume $\mb{h}_{i{q}_i}^H\mb{w}_{i{q}_i}$ to be real. 
The constraint then can be transformed into a SOC form \cite{Wiesel06}, which is convex.
However, since we now have the summation of multiple terms $\sum_{q \in \mc{Q}_i} \big|\mb{h}_{iq}^H\mb{w}_{iq}\big|^2$, with $|\mc{Q}_i| > 1$, 
there is no known method to transform the nonconvex quadratic constraint \eqref{quadratic} into a convex form, \emph{e.g.}, SOC constraint.
Thus, in order to devise an optimal solution to problem $\mc{P}_1'$, 
we rely on the nonconvex QCQP framework presented in Section \ref{QCQP-Sect}. 
Interestingly, it will be shown the
optimal solution to problem $\mc{P}'_1$ indeed meets the conditions given in Corollary \ref{coro-2}.

\section{Nonconvex Quadratic Constrained Quadratic Programming} \label{QCQP-Sect}
This section presents a brief background on nonconvex QCQP and exposes relevant
properties on strong duality of nonconvex QCQP. 
We consider a generic nonconvex QCQP as follows:
\vspace{-0.1cm}
\begin{eqnarray}\label{QCQP-1}
\mc{QCQP}\,:\quad&\underset{\mb{x}\in\mathbb{C}^N}{\minimize}&\;f_0(\mb{x}) \\
&\st&\; f_i(\mb{x}) \leq 0,\; i=1,\ldots,L,\nonumber
\end{eqnarray}
where $f_i(\mb{x}), i=0,\ldots,L$ are quadratic, but not necessarily convex, functions on $\mb{x}\in\mathbb{C}^N$.
The Lagrangian of problem \eqref{QCQP-1} is given as
\vspace{-0.1cm}
\begin{equation}
\mc{L}(\mb{x},\bs{\lambda}) = f_0(\mb{x}) + \sum_{i=1}^L \lambda_if_i(\mb{x}),
\end{equation}
where $\bs{\lambda} \triangleq [\lambda_1,\ldots,\lambda_L]^T$ and $\lambda_i \geq 0$ is
the Lagrangian multiplier associated with constraint $f_i(\mb{x})\leq 0,i=1,\ldots,L$. The dual function is
then given by
\begin{equation}
g(\bs{\lambda}) = \inf_{\mb{x}} \mc{L}(\mb{x},\bs{\lambda}).
\end{equation}
By nature, the dual function $g(\bs{\lambda})$ is concave on $\bs{\lambda}\in\mathbb{R}^L_+$ \cite{Boyd04}.
Let $p^{\star}$ be the optimal value of problem $\mc{QCQP}$ and $d^{\star}$ be the optimal value
of the dual problem $\maximize_{\bs{\lambda} \geq 0} g(\bs{\lambda})$. By definition \cite{Boyd04},
one has
\vspace{-0.1cm}
\begin{eqnarray}
&&p^{\star} = \min_{\mb{x}}\; \sup_{\bs{\lambda} \geq \mb{0}} \mc{L}(\mb{x},\bs{\lambda}), \\
\textrm{and }&& d^{\star} = \max_{\bs{\lambda} \geq \mb{0}}\; \inf_{\mb{x}} \mc{L}(\mb{x},\bs{\lambda}). \label{dual-prob}
\end{eqnarray}

Weak duality dictates that $d^{\star} \leq p^{\star}$ and the difference
$p^{\star} - d^{\star}$ is called the duality gap (cf. Section 5 in \cite{Boyd04}). If strong duality holds, \emph{i.e.},
zero duality gap with $d^{\star} = p^{\star}$, the optimal solution of the 
primal problem can be found through the dual problem as in \eqref{dual-prob}.
While strong duality holds for any convex optimization problem with Slater's condition qualification,
strong duality also obtains for nonconvex problems on rare occasions \cite{Boyd04}.
In any case of having strong duality, 
a saddle point $({\mb{x}}^{\star},{\bs{\lambda}}^{\star})$ for function $\mc{L}(\mb{x},\bs{\lambda})$, defined as
\begin{eqnarray} \label{saddle-point}
\mc{L}({\mb{x}}^{\star},\bs{\lambda}) \!\leq\! \mc{L}({\mb{x}}^{\star},{\bs{\lambda}}^{\star}) 
\!\leq\! \mc{L}(\mb{x},{\bs{\lambda}}^{\star}),\forall \mb{x}\in\mathbb{C}^N,\forall \bs{\lambda} \in \mathbb{R}^L_+,\;\;
\end{eqnarray}
must exist. The following property, presented in Section 5.4 of \cite{Boyd04}, 
underlines the connection between the existence of a saddle point
for $\mc{L}(\mb{x},\bs{\lambda})$ and strong duality.

\begin{property}
	If the function $\mc{L}(\mb{x},\bs{\lambda})$ possesses a saddle point $({\mb{x}}^{\star},{\bs{\lambda}}^{\star})$ on
	$\mathbb{C}^N \times \mathbb{R}_+^{L}$, then strong duality $d^{\star} = p^{\star}$ holds. Conversely,
	if $d^{\star}$ is finite with ${\bs{\lambda}}^{\star} = \arg \max_{\bs{\lambda}\geq \mb{0}} g(\bs{\lambda})$,
	and the original problem has an optimal solution at ${\mb{x}}^{\star}$, then $({\mb{x}}^{\star},{\bs{\lambda}}^{\star})$
	is a saddle point of $\mc{L}(\mb{x},\bs{\lambda})$.
\end{property}

The following property concerning the conditions on the existence of a saddle point 
has been presented in \cite{HoangTuy13} and its proof was partially sketched in Page 1063 of the work.

\begin{property} \label{condition}
	The existence of a saddle point of $\mc{L}(\mb{x},\bs{\lambda})$ on $\mathbb{C}^N \times \mathbb{R}_+^{L}$
	is equivalent to the following condition: there exists ${\bs{\lambda}}^{\star}$ such that the function
	$\mc{L}(\mb{x},{\bs{\lambda}}^{\star})$ is convex on $\mathbb{C}^{N}$ and has a minimizer ${\mb{x}}^{\star}$ on
	$\mathbb{C}^{N}$ satisfying ${\lambda}_i^{\star}f_i(\mb{x}^{\star}) = 0, f_i(\mb{x}^{\star}) \leq 0, i=1,\ldots,L$.
\end{property}

Thus, in order to prove strong duality in a nonconvex QCQP problem and obtain its optimal solution
via its Lagrangian dual problem, it suffices to show
that the condition given in Property \ref{condition} is fulfilled \cite{HoangTuy13}. 
Strong duality in nonconvex QCQP is also guaranteed under the following property,
which is presented as Theorem 6 in \cite{HoangTuy13}.

\begin{property} \label{theorem-3} Assume that the concave dual function 
	$g(\bs{\lambda}) = \inf_{\mb{x}} \mc{L}(\mb{x},\bs{\lambda})$ attains its maximum at a point $\bs{\lambda}^{\star} \in\mathbb{R}^L_+$.
	If $\mc{L}(\mb{x},\bs{\lambda}^{\star})$ is strictly convex on $\mathbb{C}^{N}$, then strong duality holds.
\end{property}
\section{QCQP Solution Approach to Problem $\mc{P}'_1$} \label{QCQP-approach}
This section presents an analytical approach to obtain an optimal solution to problem $\mc{P}'_1$. It is noted
that problem $\mc{P}'_1$ is a nonconvex QCQP, which is NP-hard in general \cite{HoangTuy13}. 
Our approach is to prove strong duality of this particular problem $\mc{P}'_1$. 
First, the Lagrangian of problem $\mc{P}'_1$ can be stated as
\hspace{-0.15cm}
\begin{eqnarray} \label{Lagrangian}
\hspace{-0.2cm}&&\mc{L}_1(\{\mb{w}_{iq}\},\bs{\lambda}) \nonumber \\
\hspace{-0.2cm}&&= \sum_{q=1}^Q w_q  \sum_{i\in\mc{K}_q}\|\mb{w}_{iq}\|^2 \nonumber \\
\hspace{-0.2cm}&&\;\;-\sum_{i=1}^K\!\lambda_i\bigg(\frac{1}{\gamma_{i}}\!\!\sum_{q\in\mc{Q}_i}\!\big|\mb{h}_{iq}^H\mb{w}_{iq}\big|^2 -
\sum\limits_{j\neq i}^K\sum\limits_{r\in\mc{Q}_j}\left|\mb{h}_{ir}^H\mb{w}_{jr}\right|^2 - \sigma^2 \bigg) \nonumber \\
\hspace{-0.2cm}&&=  \sum_{i=1}^K\lambda_i\sigma^2 \nonumber \\
\hspace{-0.2cm}&&\;\;+ \sum_{i=1}^K\!\sum_{q\in\mc{Q}_i}\!\mb{w}_{iq}^H\bigg( w_q\mb{I}\! -\! \frac{\lambda_i}{\gamma_i}\mb{h}_{iq}\mb{h}_{iq}^H
\!+\! \sum_{j\neq i}^K\!\lambda_j\mb{h}_{jq}\mb{h}_{jq}^H\bigg)\mb{w}_{iq}.
\end{eqnarray}
The dual function is then given by $g_1(\bs{\lambda}) = \min_{\{\mb{w}_{iq}\}}\mc{L}_1(\{\mb{w}_{iq}\},\bs{\lambda})$.
Clearly, if any matrix $w_q\mb{I} - \frac{\lambda_i}{\gamma_i}\mb{h}_{iq}\mb{h}_{iq}^H
\!+\! \sum_{j\neq i}^K\lambda_j\mb{h}_{jq}\mb{h}_{jq}^H$ is not positive semi-definite,
it is possible to find $\mb{w}_{iq}$ to make $g_1(\bs{\lambda})$ unbounded below. Thus,
the dual problem is given by
\hspace{-0.15cm}
\begin{eqnarray} \label{dual-prob-2}
&\underset{\bs{\lambda}\geq \mb{0}}{\maximize}&\; \sum_{i=1}^K\lambda_i \sigma^2\\
&\st&\;  w_q\mb{I} + \sum_{j\neq i}^K\lambda_j\mb{h}_{jq}\mb{h}_{jq}^H \succeq \frac{\lambda_i}{\gamma_i}\mb{h}_{iq}\mb{h}_{iq}^H,
\forall q \in\mc{Q}_i,\forall i.\nonumber
\end{eqnarray}

\emph{Remark 2:} The dual problem is an SDP and a convex problem by nature. Its optimal solution
can be easily obtained by the interior point method or standard SDP solvers, such as \texttt{cvx} \cite{CVX}.
However, a closer look on the dual problem \eqref{dual-prob-2} can analytically establish an optimal solution
to problem $\mc{P}'_1$ as well as its feasibility. Note that the dual problem \eqref{dual-prob-2} is
always feasible  (for instance, $\lambda_i=0,\forall i$ satisfies all the constraints). However, its feasibility
does not necessarily indicate the feasibility of the primal problem $\mc{P}'_1$.
It may happen that $\lambda_i \rightarrow \infty$ at optimality and all constraints in \eqref{dual-prob-2} are still satisfied, \emph{i.e.},
the dual problem is unbounded above. In this case, the primal problem $\mc{P}'_1$ is infeasible thanks to the weak duality properties \cite{Boyd04}.

We now focus on the case where the optimal value of the dual problem \eqref{dual-prob-2} is finite, \emph{i.e.},
the primal problem $\mc{P}'_1$ is feasible.

\begin{theorem}\label{theorem-important}
If the nonconvex QCQP $\mc{P}'_1$ is feasible, then strong duality holds.
\end{theorem}
\begin{IEEEproof}
Denote $\bs{\lambda}^{\star}$ as the optimal solution of the dual problem. At $\bs{\lambda}^{\star}$, 
the function $\mc{L}_1(\{\mb{w}_{iq}\},\bs{\lambda}^{\star})$ is convex in $\{\mb{w}_{iq}\}$.
Thus, in order to satisfy the conditions in Property \ref{condition} as presented in Section \ref{QCQP-Sect}, 
it is left to find $\{\mb{w}_{iq}^{\star}\} 
\in \arg\min_{\{\mb{w}_{iq}\}} \mc{L}(\{\mb{w}_{iq}\},\bs{\lambda}^{\star})$ such that $\{\mb{w}_{iq}^{\star}\}$ 
is feasible to problem $\mc{P}'_1$ and moreover
\hspace{-0.1cm}
\begin{equation} \label{optimality}
\lambda_i^{\star}\!\!\left(\!\frac{1}{\gamma_{i}}\!\!\sum_{q\in\mc{Q}_i}\!\!\big|\mb{h}_{iq}^H\mb{w}_{iq}^{\star}\big|^2 \
\!-\! \sum\limits_{j\neq i}^K\!\sum\limits_{r\in\mc{Q}_j}\!\!\left|\mb{h}_{ir}^H\mb{w}_{jr}^{\star}\right|^2\! - 
\! \sigma^2\!\right)\!\! = 0,\forall i.
\end{equation}

Consider the set of constraints related to MS-$i$ with optimal $\bs{\lambda}^{\star}$
in the dual problem \eqref{dual-prob-2}. 
Suppose that
\hspace{-0.1cm}
\begin{equation}
w_q\mb{I} + \sum_{j\neq i}^K\lambda^{\star}_j\mb{h}_{jq}\mb{h}_{jq}^H \succ
\frac{\lambda^{\star}_i}{\gamma_i}\mb{h}_{iq}\mb{h}_{iq}^H,\forall q\in\mc{Q}_i.
\end{equation}
We can increase $\lambda_i^{\star}$ to some value $\hat{\lambda}_i$ such that
\hspace{-0.15cm}
\begin{equation}
w_{q}\mb{I} + \sum_{j\neq i}^K\lambda^{\star}_j\mb{h}_{jq}\mb{h}_{jq}^H \succeq
\frac{\hat{\lambda}_i}{\gamma_i}\mb{h}_{iq}\mb{h}_{iq}^H,\forall q\in\mc{Q}_i.
\end{equation}
By setting $\hat{\lambda}_j = \lambda_j^{\star}, \forall j\neq i$, one yields a feasible solution
$\hat{\bs{\lambda}} = [\hat{\lambda}_1,\ldots,\hat{\lambda}_K]^T$ that improves the
objective function of problem \eqref{dual-prob-2}, \emph{i.e.}, $\sum_{i=1}^K\hat{\lambda}_i\sigma^2 > \sum_{i=1}^K{\lambda}_i^{\star}\sigma^2$.
Thus, $\bs{\lambda}^{\star}$ cannot be an optimal solution of problem $\eqref{dual-prob-2}$ by contradiction.
As a result, there must exist a non-empty subset $\hat{\mc{Q}}_i \subset \mc{Q}_i$, such that
\hspace{-0.15cm}
\begin{equation} \label{singular}
w_{q}\mb{I} + \sum_{j\neq i}^K\lambda_j^{\star}\mb{h}_{jq}\mb{h}_{jq}^H \nsucc
\frac{\lambda_i^{\star}}{\gamma_i}\mb{h}_{iq}\mb{h}_{iq}^H, \forall q \in \hat{\mc{Q}}_i.
\end{equation}
Otherwise, $\lambda_i^{\star}$ can be further increased. 
Due to the above set of inequalities (with $\nsucc$), 
$\lambda_i^{\star}, \forall i$ must be positive.
For $q \in \mc{Q}_i \backslash \hat{\mc{Q}}_i$,
strict inequality applies, \emph{i.e.}, 
\hspace{-0.15cm}
\begin{equation} \label{strict}
w_q\mb{I} + \sum_{j\neq i}^K\lambda^{\star}_j\mb{h}_{jq}\mb{h}_{jq}^H \succ
\frac{\lambda^{\star}_i}{\gamma_i}\mb{h}_{iq}\mb{h}_{iq}^H,\forall q\in\mc{Q}_i \backslash \hat{\mc{Q}}_i.
\end{equation}

Since strict inequality \eqref{strict} is enforced $\forall q\in\mc{Q}_i \backslash \hat{\mc{Q}}_i$,
the corresponding beamforming vector $\mb{w}^{\star}_{iq}$ must be set 
to all-$0$ vector in order to have the Lagrangian function  \eqref{Lagrangian} minimized.
On the other hand, since inequality \eqref{singular} happens for $q \in \hat{\mc{Q}}_i$, there exists
an eigenvector $\hat{\mb{w}}_{iq} \in\mathbb{C}^M, \|\hat{\mb{w}}_{iq}\| = 1$ corresponding to the $0$ eigenvalue, such that
\begin{equation} \label{dual-equations}
\hat{\mb{w}}_{iq}^H\Bigg(\!w_{q}\mb{I} +\! \sum_{j\neq i}^K\lambda_j^{\star}\mb{h}_{jq}\mb{h}_{jq}^H\! -
\frac{\lambda_i^{\star}}{\gamma_i}\mb{h}_{iq}\mb{h}_{iq}^H\Bigg)\hat{\mb{w}}_{iq}\! = 0, \forall i, \forall q \in \hat{\mc{Q}}_i.
\end{equation}

To minimize the Lagrangian $\mc{L}_1(\{\mb{w}_{iq}\},\bs{\lambda}^{\star})$, $\mb{w}_{iq}^{\star}$ 
can be chosen as a scaled version of $\hat{\mb{w}}_{iq}$.
For each MS, say MS-$i$, a BS indexed as $q_i \in \hat{\mc{Q}}_i$ is randomly chosen 
and the corresponding beamforming vector $\mb{w}^{\star}_{iq_i}$ is set as $\sqrt{\delta_i}\hat{\mb{w}}_{iq_i}$,
where the scaling coefficient $\delta_i > 0$ will be determined shortly. For all BSs, $q \in \hat{\mc{Q}}_i$, $q\neq q_i$,
$\mb{w}^{\star}_{iq}$ is purposely set at $\mb{0}$. Thus, we obtain a set of beamforming vector 
$\{\mb{w}_{iq}^{\star}\}_{\forall q\in\mc{Q}_i} \in \arg\min_{\{\mb{w}_{iq}\}} \mc{L}(\{\mb{w}_{iq}\},\bs{\lambda}^{\star})$.
The next step is to determine $\delta_i$'s such that $\mb{w}^{\star}_{iq_i}$'s satisfies condition \eqref{optimality}.

Note that $\lambda_i^{\star} > 0,\forall i$ and $\mb{w}^{\star}_{iq} = \mb{0}, \forall q\neq q_i$.
By substituting $\mb{w}^{\star}_{iq_i} = \sqrt{\delta_i}\hat{\mb{w}}_{iq_i}$ into \eqref{optimality},
we obtain a set of equations 
\vspace{-0.1cm}
\begin{eqnarray} \label{set-equations}
\!\!\!\!\!\frac{\delta_i}{\gamma_{i}}\big|\mb{h}_{iq_i}^H\hat{\mb{w}}_{iq_i}\big|^2 -
\delta_j\sum\limits_{j\neq i}^K\big|\mb{h}_{iq_j}^H\hat{\mb{w}}_{jq_j}\big|^2 = \sigma^2, i=1,\ldots,K,
\end{eqnarray}
Equivalently, $\mb{G}\bs{\delta} = \mb{1}\sigma^2$,
where $\bs{\delta} = [\delta_1,\ldots,\delta_K]^T$ and
$\mb{G} \in \mathbb{R}^{K\times K}$ is defined as $[\mb{G}]_{i,i} = (1/\gamma_i)\big|\mb{h}_{iq_i}^H\hat{\mb{w}}_{iq_i}\big|^2$
and $[\mb{G}]_{i,j} = -\big|\mb{h}_{iq_j}^H\hat{\mb{w}}_{jq_j}\big|^2$.

It is noted that the set of equations in \eqref{dual-equations}
can be cast as
\vspace{-0.1cm}
\begin{equation}
\mb{G}^T\bs{\lambda}^{\star} = [w_{q_1},\ldots,w_{q_K}]^T > \mb{0}.
\end{equation}
Since $\mb{G}^T$ is a $\mb{Z}$-matrix and there exists $\bs{\lambda}^{\star} > 0$ such that
$\mb{G}^T\bs{\lambda}^{\star} > \mb{0}$, $\mb{G}^T$ is an $\mb{M}$-matrix by its characterization
(Condition I$_{28}$, Theorem 6.2.3 in \cite{Berman79}).\footnote{
	A square matrix $\mb{X}$ is a $\mb{Z}$-matrix if all its off-diagonal elements are nonpositive.
	A square matrix $\mb{X}$ is a $\mb{P}$-matrix if all its principle minors are positive.
	A square matrix that is both a $\mb{Z}$-matrix and a $\mb{P}$-matrix is called a $\mb{M}$-matrix \cite{Berman79,Cottle92}.} 
Thus, $\mb{G}$, also
an $\mb{M}$-matrix, is invertible and its inverse is a positive matrix \cite{Berman79}.
As a result, $\bs{\delta} > \mb{0}$ can be determined by
\hspace{-0.1cm}
\begin{equation}\label{delta}
\bs{\delta} = \mb{G}^{-1}\mb{1}\sigma^2.
\end{equation}
Since $\{\mb{w}_{iq_i}^{\star} = \sqrt{\delta_i}\hat{\mb{w}}_{iq_i}\}_{\forall i}$ now satisfies the set of equations \eqref{dual-equations},
we yield a feasible solution to the problem $\mc{P}_1'$ where each constraint is met with equality.
The qualification of condition \eqref{optimality} by
$\{\mb{w}_{iq_i}^{\star}\}$ then guarantees the satisfaction of all conditions in Property \ref{condition}.
Strong duality for problem $\mc{P}_1'$ then follows. 
Furthermore, $\{\mb{w}_{iq_i}^{\star}\}$ must be a globally optimal solution of the nonconvex problem
$\mc{P}'_1$.
\end{IEEEproof}

We now relate the optimal solution $\{\mb{w}_{iq_i}^{\star}\}$ of problem $\mc{P}_1'$ to the
original mixed-integer problem $\mc{P}_1$ as follows.

\begin{proposition}
The obtained optimal solution $\{\mb{w}_{iq_i}^{\star}\}$ of problem $\mc{P}_1'$
is also optimal to the original mixed-integer problem $\mc{P}_1$. Furthermore, $q_i$ indicates 
an optimal BS association for MS-$i$.
\end{proposition}
\begin{IEEEproof}
In solving problem $\mc{P}_1'$, we derived an optimal solution where $\mb{w}_{iq_i}^{\star} \neq \mb{0}$ 
and $\mb{w}_{iq}^{\star} = \mb{0},\forall q \neq q_i$. Thus, Corollary \ref{coro-2} is applicable.
The optimality of $\{\mb{w}_{iq_i}^{\star}\}$ to problem $\mc{P}_1$ and the association of MS-$i$ to
BS-$q_i$ follow.
\end{IEEEproof}

Through numerous numerical simulations, we observe that inequality \eqref{singular} is met at only one BS 
in the set $\mc{Q}_i$, \emph{i.e.}, $|\hat{\mc{Q}}_i| = 1$, except for the extremely rare cases where
the channels from two BSs are exactly symmetric or two BSs are co-located. We address the cases when 
$|\hat{\mc{Q}}_i| > 1$ in the following proposition.
\begin{proposition}
	If $|\hat{\mc{Q}}_i| > 1$, MS-$i$ can be associated to either one of the BSs in the set $\mc{\hat{Q}}_i$
	without affecting minimum weighted sum transmit power across the BSs.
\end{proposition}
\begin{IEEEproof}
If $|\hat{\mc{Q}}_i| > 1$, we can first select any BS, say $q_i \in \hat{\mc{Q}}_i$, such
that the corresponding beamformer $\mb{w}_{q_i}^{\star}$ is set to be non-zero and $\mb{w}_{q}^{\star} = \mb{0},\forall q\neq q_i$.
The derivation steps \eqref{dual-equations}--\eqref{delta} can be sequentially applied 
to determine the scaling factor and the beamformers $\{\mb{w}_{iq}^{\star}\}$ for all the users.
Interestingly, different association schemes (with $q_i \in \mc{\hat{Q}}_i$ and their corresponding beamforming designs)
might yield different globally optimal solutions to problem $\mc{P}_1'$ with the same optimal value. The reason for this result is because 
the obtained solutions $(\mb{w}_{iq}^{\star},\bs{\lambda}^{\star})$ will satisfy the set of equations 
\eqref{optimality} and other conditions in Property \ref{condition} to be globally optimal.
In addition, different BS assignments for MS-$i$ in $\hat{\mc{Q}}_i$ will also yield the same minimum
weighted sum power across the BSs, which must equal to the optimal value of the dual problem \eqref{dual-prob-2},
$\sum_{i=1}^K\lambda_i^{\star}\sigma^2$.
In spite of that, individual transmit powers at the BSs might not be the same with different BS assignments for MS-$i$.
\end{IEEEproof}

Since the $\mc{P}_1$ can be optimally solved, any Pareto-optimal solution of the problem can 
be obtained by properly adjusting the weight factor $w_q$'s in the objective function.
\vspace{-0.1cm}
\section{Interpretation via Uplink-Downlink Duality} \label{Uplink-Downlink-approach}
In the previous section, we have presented an analytical approach to solve problem $\mc{P}'_1$
via its Lagrangian dual problem. In this section, we provide an alternative approach for solving problem $\mc{P}'_1$
via the well-known uplink-downlink duality. It will be shown shortly that the Lagrangian
dual problem \eqref{dual-prob-2} is indeed the power minimization problem with SINR constraints
in the uplink. We note that uplink-downlink duality is a powerful tool
which has been studied in different contexts of multicell beamforming designs \cite{WeiYu10,Huang-Ottersten-TSP2011,He-Huang-TWC2012,Huang-Ottersten-TWC2012,Huang-Tan-Rao-TWC2013}. Fixed-point iterative
algorithms were proposed to find the corresponding optimal beamforming solutions \cite{WeiYu10,Huang-Ottersten-TSP2011,He-Huang-TWC2012,Huang-Ottersten-TWC2012}. Herein,
we show that uplink-downlink duality is also applicable to the joint BS association
and beamforming design problem under consideration. We then propose an iterative fixed-point algorithm to effectively solve the problem.
\vspace{-0.15cm}
\subsection{Dual Uplink System Model}
We consider the dual uplink system with the same setting as in Section \ref{sect-2}. Specifically,
the dual uplink system is derived from the downlink system by transposing the channel matrices
and by interchanging the input and the output vectors. In addition, the noise at each BS, say BS-$q$,
is assumed to be zero mean AWGN with the covariance matrix $\sigma^2w_q\mb{I}$.
Herein, the single-antenna MS-$q$ is transmitting at power $\hat{p}_i$ and $\mb{h}_{iq}$ is
the uplink channel from MS-$i$ to BS-$q$. If MS-$i$ is associated with BS-$q$ where $q\in\mc{Q}_i$,
the BS then applies the receive beamforming vector $\hat{\mb{w}}_{iq}$ to decode MS-$i$'s signal.
In the considered uplink system, the BS association is performed by selecting a BS in $\mc{Q}_i$ such that
MS-$i$ needs to transmit at minimum power to obtain the SINR target $\gamma_i$ at the very BS. Thus,
the design objective now is to jointly optimize the power allocation $\hat{p}_i$'s,
the receive beamforming vector $\hat{\mb{w}}_{iq},q\in\mc{Q}_i, \forall i$, and the BS association
to satisfy the set of SINR constraints $\gamma_i$'s. The joint uplink optimization
can be stated as
\vspace{-0.1cm}
\begin{eqnarray} \label{uplink}
\hspace{-0.2cm}&\underset{\hat{p}_1,\ldots,\hat{p}_K,\{\hat{\mb{w}}_{iq}\}}{\minimize}&\! \sum_{i=1}^K\hat{p}_i\\
\hspace{-0.2cm}&\st&\! \max_{q\in\mc{Q}_i} \frac{\hat{p}_i\big|\hat{\mb{w}}_{iq}^H\mb{h}_{iq}\big|^2}
{\sum\limits_{j\neq i}^K\hat{p}_j\big|\hat{\mb{w}}_{iq}^H\mb{h}_{jq}\big|^2
+ \sigma^2w_q\hat{\mb{w}}_{iq}^H\hat{\mb{w}}_{iq}}\geq \gamma_i, \forall i. \nonumber
\end{eqnarray}

We now underline the connection between the downlink problem $\mc{P}'_1$ and the dual uplink problem \eqref{uplink}.
\begin{proposition}
The optimal downlink beamforming problem $\mc{P}'_1$ can be solved via a dual uplink problem in which
the SINR constraints remain the same. Specifically, the Lagrangian dual problem \eqref{dual-prob-2} of
problem $\mc{P}'_1$ is the following problem
\vspace{-0.1cm}
\begin{eqnarray} \label{uplink-virtual}
\hspace{-0.4cm}&\underset{\lambda_1,\ldots,\lambda_K,\{\hat{\mb{w}}_{iq}\}}{\minimize}&\! \sum_{i=1}^K\lambda_i \sigma^2\\
\hspace{-0.4cm}&\st&\! \max_{q\in\mc{Q}_i} \frac{\lambda_i\sigma^2\big|\hat{\mb{w}}_{iq}^H\mb{h}_{iq}\big|^2}
{\sum\limits_{j\neq i}^K\!\lambda_j\sigma^2\big|\hat{\mb{w}}_{iq}^H\mb{h}_{jq}\big|^2
\!+ \sigma^2w_q\hat{\mb{w}}_{iq}^H\hat{\mb{w}}_{iq}}\!\geq\! \gamma_i, \forall i, \nonumber
\end{eqnarray}
where $\hat{p}_i = \lambda_i\sigma^2$ is dual uplink power of MS-$i$.
If the dual uplink problem \eqref{uplink-virtual} is feasible, its optimal solution is also optimal
to the Lagrangian dual problem \eqref{dual-prob-2}. Otherwise, the Lagrangian dual problem \eqref{dual-prob-2} is
unbounded above.
\end{proposition}
\begin{IEEEproof}
For given uplink power allocation $\hat{p}_i = \lambda_i\sigma^2$, 
the optimal receive beamforming vector
at BS-$q,q\in\mc{Q}_i$ is the minimum mean-squared error (MMSE) receiver
\vspace{-0.1cm}
\begin{eqnarray}\label{MMSE-receiver}
\hat{\mb{w}}_{iq} = \Bigg(\sum_{j=1}^K\lambda_j\mb{h}_{jq}\mb{h}_{jq}^H + w_q\mb{I}\Bigg)^{-1}\mb{h}_{iq}.
\end{eqnarray}
By substituting the above MMSE receiver $\hat{\mb{w}}_{iq}$, the SINR constraint for MS-$i$ in \eqref{uplink-virtual} becomes
\begin{equation} \label{const-uplink}
\!\!\lambda_i\!\left(\!1 + \frac{1}{\gamma_i}\right)\!\cdot\max_{q\in\mc{Q}_i}\mb{h}_{iq}^H\Bigg(
\sum_{j=1}^K\lambda_j\mb{h}_{jq}\mb{h}_{jq}^H + w_q\mb{I}\!\Bigg)^{-1}\!\!\mb{h}_{iq} \geq 1.\!\!
\end{equation}
Note that the above set of constraints for $i=1,\ldots,K$ may constitute an empty set, which then renders the dual uplink problem infeasible.
However, if the dual uplink problem \eqref{uplink-virtual} is feasible, at optimality the set of inequality
constraints \eqref{const-uplink} must meet at equality, \emph{i.e.},
\vspace{-0.1cm}
\begin{equation} \label{equal-const}
\!\!\lambda_i\!\left(\!1\! + \frac{1}{\gamma_i}\right)\!\cdot\max_{q\in\mc{Q}_i}\mb{h}_{iq}^H
\Bigg(\sum_{j=1}^K\lambda_j\mb{h}_{jq}\mb{h}_{jq}^H + w_q\mb{I}\!\Bigg)^{-1}\!\!\mb{h}_{iq} = 1, \forall i.
\end{equation}

Thanks to Lemma 1 in \cite{WeiYu07} as provided following this proof, 
the constraint in the Lagrangian dual problem \eqref{dual-prob-2}
can be recast as
\begin{eqnarray}
\!\lambda_i\left(\!1\! +\! \frac{1}{\gamma_i}\right)\mb{h}_{iq}^H\Bigg(\!
\sum_{j=1}^K\lambda_j\mb{h}_{jq}\mb{h}_{jq}^H + w_q\mb{I}\!\Bigg)^{-1}\!\!\mb{h}_{iq}\leq 1,
\forall q \in\mc{Q}_i,&&\nonumber
\end{eqnarray}
or equivalently,
\vspace{-0.1cm}
\begin{equation} \label{const-dual}
\lambda_i\!\left(\!1 + \frac{1}{\gamma_i}\right)\!\cdot\max_{q\in\mc{Q}_i}\mb{h}_{iq}^H\Bigg(
\sum_{j=1}^K\lambda_j\mb{h}_{jq}\mb{h}_{jq}^H + w_q\mb{I}\!\Bigg)^{-1}\!\!\mb{h}_{iq} \leq 1.
\end{equation}

Unlike the dual uplink problem \eqref{uplink-virtual}, the Lagrangian dual problem \eqref{dual-prob-2} is always
feasible, thanks to its nonempty constraint set. In case of having a finite optimal value,
it is clear that at optimality the set of inequality constraints \eqref{const-dual} must be met at equality, as
given in \eqref{equal-const}.
Thus, the power minimization problem \eqref{uplink-virtual} of $\sum_{i=1}^K\lambda_i\sigma^2$
with minimum SINR constraints \eqref{const-uplink} and the power maximization problem \eqref{dual-prob-2}
of $\sum_{i=1}^K\lambda_i\sigma^2$ with maximum SINR constraints in \eqref{const-dual}
are equivalent since $\lambda_i$'s in both problems are the fixed point of the equations \eqref{equal-const}.
It will be shown shortly that this fixed point is unique if it exists. In that case, the fixed point is the optimal solution
for both problems. If a fixed point does not exist, the dual uplink problem \eqref{uplink-virtual}
is not feasible and equivalently the Lagrangian dual problem \eqref{dual-prob-2} is unbounded above.
\end{IEEEproof}
For completeness, Lemma 1 in \cite{WeiYu07} is presented as follows: \emph{``Let $\mb{A}$ be an $n\times n$ 
	symmetric positive semidefinite matrix and $\mb{b}$ be an $n\times 1$ vector.
Then, $\mb{A} \succeq \mb{b}\mb{b}^H$ if and only if $\mb{b}^H\mb{A}^{-1}\mb{b} \leq 1$.''}

\vspace{-0.1cm}
\subsection{An Iterative Algorithm for Solving Problem $\mc{P}'_1$} \label{Uplink-Downlink-iteration} 
Having established the equivalence between the Lagrangian dual problem \eqref{dual-prob-2}
and the dual uplink problem \eqref{uplink-virtual}, this section focuses on obtaining the solution
to both problems by finding the fixed point to the set of equations \eqref{equal-const}.
By rearranging \eqref{equal-const} into a fixed point iteration, one has
\begin{eqnarray} \label{iteration}
\lambda_i^{(n+1)} &=& \min_{q\in\mc{Q}_i} f_{iq}\big(\bs{\lambda}^{(n)}\big),
\end{eqnarray}
where $f_{iq}(\bs{\lambda})$ is defined as
\begin{equation}
f_{iq}(\bs{\lambda}) = \frac{\gamma_i}{1+\gamma_i} \cdot
\frac{1}{\mb{h}_{iq}^H\mb{\Sigma}_q^{-1}\!\mb{h}_{iq}}.
\end{equation}
and $\mb{\Sigma}_q = \!\sum_{j=1}^K\lambda_j\mb{h}_{jq}\mb{h}_{jq}^H + w_q\mb{I}$. 

\begin{proposition} \label{prop-3}
If a fixed point of \eqref{equal-const} exists, it is unique and
the iterative function evaluation \eqref{iteration} converges geometrically fast to the fixed point.
\end{proposition}
\begin{IEEEproof}
It is proven in \cite{Wiesel06,WeiYu07} that $f_{iq}(\bs{\lambda})$ satisfies the three properties (positivity, monotonicity, and
scalability) to be a standard function. Moreover, the point-wise minimum of a set of standard function, \emph{i.e.},
$\min_{q\in\mc{Q}_i}f_{iq}(\bs{\lambda})$, is also a standard function \cite{Yates95}. Thus,
the iteration \eqref{iteration} converges geometrically fast to the fixed point, if it exists.
\end{IEEEproof}

The iteration \eqref{iteration} accounts for the first step of the iterative algorithm to solve problem $\mc{P}_1'$.
The second step is to find the optimal receive beamformer $\hat{\mb{w}}_{iq_i}$, where $q_i$ is the BS association 
with MS-$i$. Then, the final step is to obtain the optimal transmit beamformer ${\mb{w}}_{iq_i}$.
We summarize these three steps in the following Algorithm \ref{algo-1}.
\begin{algorithm}
\small
Initialize $\lambda_i >0$, $\forall i$\;
\SetAlgoNoLine
\Repeat{convergence to $\bs{\lambda}^{\star}$}{
Update $\lambda_i$ as given in equation \eqref{iteration},}
Set $q_i = \arg\min_{q\in\mc{Q}_i}f_{iq}(\bs{\lambda}^{\star}),\forall i$\;
Find the receive beamformer $\hat{\mb{w}}_{q_i}$ as given in equation \eqref{MMSE-receiver}\;
Find the transmit beamformer $\mb{w}_{iq_i}^{\star} = \sqrt{\delta_i}\hat{\mb{w}}_{q_i}$, where $\delta_i$ is given by
\eqref{delta}\;
\caption{Iterative Algorithm for Minimizing The Weighted Sum Transmit Power}
\label{algo-1}
\end{algorithm}
\normalsize
 
\vspace{-0.5cm}
\subsection{Distributed Implementation}
An interesting development from the above uplink-downlink duality interpretation is that all the three
steps in the iterative algorithm proposed in the previous section can be implemented distributively. Herein, it is assumed that
the system is operating in the time division duplex (TDD) mode where the uplink and downlink channels are reciprocal. 
It is also assumed that the weight $w_q$ is known at BS-$q$.

In the first step, the iteration \eqref{iteration} on the uplink power $\lambda_i$ involves 
only its channel vectors $\mb{h}_{iq}$'s and the matrices $\mb{\Sigma}_q$'s obtained
from the BSs in $\mc{Q}_i$. With known $w_q$, BS-$q$ can compensate the background noise to $w_q\sigma^2$. 
Then $\mb{\Sigma}_q$, as the covariance matrix of the total received
signal at BS-$q$ in the uplink, can be estimated locally by the BS. Thus, the transmit powers $\lambda_i$'s
can be updated as in \eqref{iteration} on a per-user basis without inter-BS or inter-user coordination. Should 
the acquisition of the channel $\mb{h}_{iq}$'s or the matrices $\mb{\Sigma}_q$'s be challenging at the MSs,
BS-$q$ can simply calculate $f_{iq} = \frac{\gamma_i}{1+\gamma_i}\cdot\frac{1}{\mb{h}_{iq}^H\mb{\Sigma}_q^{-1}\!\mb{h}_{iq}}$ for $i\in\mc{K}_q$
as the required transmit power at MS-$i$ to obtain its target SINR $\gamma_i$ at the very BS. Subsequently, $f_{iq}$'s are passed
to MS-$i$, who will choose the lowest uplink power $\lambda_i = \min_{q\in\mc{Q}_i}f_{iq}$. The BS that can achieve
the SINR $\gamma_i$ with the uplink power $\lambda_i$ is then the one associated with MS-$i$.
Thanks to Proposition \ref{prop-3}, these iterative steps always converge to a fixed point if it exists.

While the second step to determine the MMSE receivers \eqref{MMSE-receiver} is straightforward at the BSs,
the final step to calculate the scaling factors $\delta_i$'s is more involved. In particular, 
although $\delta_i$'s are found as in \eqref{delta}, this matrix inversion process requires centralized implementation.
On the other hand, finding $\delta_i$'s is equivalent to the downlink power control problem 
for achieving a set target SINRs $\gamma_i$'s. One solution approach is the 
Foschini-Miljanic's algorithm where the optimal downlink powers can be found iteratively in a fully
distributed manner using per-user power updates \cite{Foschini93}.

\emph{Remark 3:} In downlink CS/CB, it is shown in \cite{WeiYu10} that the optimal downlink beamforming
can be obtained even if each BS only knows the CSI to its connected MSs by exploiting channel reciprocality.
Via the distributed implementation presented in this section, we show that the optimal
BS association and beamforming design in the DPS mode can be devised 
if each BS only knows the CSI to the MSs in its serving user set, \emph{i.e.},
BS-$q$ needs the CSI $\mb{h}_{iq}$ to the MSs in $\mc{K}_q$. By exploiting channel reciprocality, BS-$q$ can listen to the
training signal from MS-$i$ in the uplink transmission for estimating $\mb{h}_{iq}$. 
The only signaling or feedback involved
is passing of the required transmit power $f_{iq}$ for MS-$i$ to connect to BS-$q$ in the uplink.
MS-$i$ is then required to make the recommendation of its selected BS $q_i$, \emph{i.e.}, $q_i = \arg\min_{q\in\mc{Q}_i} f_{iq}$.
This selection recommendation by the MSs is consistent with the CoMP implementation in the LTE Release 11 \cite{3GPPTR36.819}.

\vspace{-0.15cm}
\section{Semidefinite Programming Relaxation}
In this section, we present the SDP relaxation approach to find an optimal solution to problem $\mc{P}'_1$.
It is well known that SDP relaxation can be successfully exploited to find the optimal multiuser
beamforming design for single-cell systems \cite{Bengt01,Wiesel06,Wiesel08}. 
To apply the SDP relaxation to the multicell system model under consideration, we first replace $\mb{w}_{iq}\mb{w}_{iq}^H$ by $\mb{X}_{iq}$
and $\mb{h}_{iq}\mb{h}_{iq}^H$ by $\mb{H}_{iq}$ and recast problem $\mc{P}'_1$ into an SDP
\vspace{-0.1cm}
\begin{eqnarray} \label{SDP}
\!\!\!\!\!&\underset{\{\mb{X}_{iq}\}}{\minimize}&\! \sum_{q=1}^Qw_q \sum_{i\in \mc{K}_q} \tr\{\mb{X}_{iq}\} \\
\!\!\!\!\!&\st&\! \frac{1}{\gamma_i}\!\!\sum_{q \in \mc{Q}_i}\!\! \tr\{\mb{H}_{iq}\mb{X}_{iq}\} \! - \!
\sum\limits_{j\neq i}^K\!\sum\limits_{r\in\mc{Q}_j}\!\!\tr\{\mb{H}_{ir}\mb{X}_{jr}\!\}\! \geq\! \sigma^2,\forall i \nonumber\\
\!\!\!\!\!&&\mb{X}_{iq} \succeq \mb{0},\; \mr{rank}\{\mb{X}_{iq}\} = 1. \nonumber
\end{eqnarray}

Since the rank constraint is nonconvex, we remove it and relax problem \eqref{SDP} into
a convex SDP. Once we have a convex SDP, the interior-point method can be applied to find its optimal solution. 
Through numerous numerical simulations, we found a similar result reported \cite{Bengt01,Wiesel08}
that a rank-$1$ solution of the SDP relaxation problem can always be found. Thus, it is possible to
retrieve $\mb{w}_{iq}$ from the obtained solution in $\mb{X}_{iq}$. 
In addition, it is even more interesting that among the optimal solution set related to a particular MS, \emph{e.g.}, $\{\mb{X}_{iq}\},q\in\mc{Q}_i$, 
there is only \emph{one} non-zero (and rank-$1$) matrix.
As a result, solving the SDP relaxation version of problem \eqref{SDP} provides the optimal
solution not only to the beamforming problem $\mc{P}'_1$, but also to the original joint BS association and beamforming
design problem $\mc{P}_1$. 

It is noted that the obtained optimal result from the SDP relaxation approach can be 
proved analytically. First, it can be shown that the Lagrangian dual of the SDP relaxation version of problem \eqref{SDP} is the same as 
problem \eqref{dual-prob-2}. Second, problem \eqref{dual-prob-2} is the dual problem of the QCQP $\mc{P}_1'$
and strong duality holds. Thus, our proposed framework via nonconvex QCQP in Section \ref{QCQP-approach} provides a rigorous 
analytical confirmation to the numerical results obtained here by the SDP relaxation approach.
Nonetheless, the drawback of the SDP relaxation approach is the complexity in solving the
relaxed version of problem \eqref{SDP} due to the expanded set of variables. Certainly, solving the convex SDP in the
Lagrangian dual problem \eqref{dual-prob-2} is much simpler.
\vspace{-0.2cm}
\section{Minimization of The Per-Base-Station Transmit Power Margin} \label{per-BS-sect}
In the optimization problem $\mc{P}_1$, the adjustment of the weight factors $w_q$'s provides a trade-off among 
the power consumptions at different BSs. In this section, we consider a practical 
scenario of minimizing the transmit power margin across the BSs, in which the weights are implicitly determined.
To jointly optimize the BS association and the beamforming design, the optimization problem can be formulated as follows:
\begin{eqnarray} \label{opt-prob-3}
\mc{P}_2\,:\quad&\underset{\{a_{iq}\},\{\mb{w}_{iq}\},\alpha}{\minimize}&\; \alpha\sum_{q=1}^Q P_q \\
&\st&\; \sum_{q \in \mc{Q}_i} a_{iq}\mr{SINR}_{iq} \geq \gamma_i,\forall i \nonumber\\
&&\; a_{iq} = \{0,1\},\forall q\in\mc{Q}_i,\forall i \nonumber \\
&&\; \sum_{q\in\mc{Q}_i} a_{iq} = 1, \forall i, \nonumber \\
&&\; \sum_{i\in\mc{K}_q}\left\|\mb{w}_{iq}\right\|^2 \leq \alpha P_q.\nonumber
\end{eqnarray}
Herein, $\alpha$ represents the margin between the transmit power of a BS, 
say BS-$q$, and its maximum power value $P_q$. By minimizing $\alpha$, the multicell system tries to balance the power
consumptions across the BSs and does not overuse any of them. This formulation is especially beneficial 
to heterogeneous multicell systems where $P_q$'s can be different by one or two orders of magnitude. The resulting optimal $\alpha^{\star}$ from problem $\mc{P}_2$ is also important to verify the compliance of individual power constraints at the BSs. 
Specifically, if $\alpha^{\star} \leq 1$, then it is feasible to find an optimal BS association and beamforming design to meet all the SINR constraints and the per-BS power constraint $\sum_{i\in\mc{{K}}_q} \|\mb{w}_{iq}\|^2 \leq P_q,\forall q$.

Similar to problem $\mc{P}_1$, problem
$\mc{P}_2$ is a difficult nonconvex mixed integer program. Thus, we take a similar approach in solving problem $\mc{P}_1$
by relaxing problem $\mc{P}_2$ into the following optimization problem:
\begin{eqnarray}\label{opt-prob-4}
\mc{P}_2'\,:\quad&\underset{\{\mb{w}_{iq}\},\alpha}{\minimize}&\; \alpha\sum_{q=1}^Q P_q \\
&\st&\; \sum_{q \in \mc{Q}_i} \frac{\big|\mb{h}_{iq}^H\mb{w}_{iq}\big|^2}{
\sum\limits_{j\neq i}^K\sum\limits_{r\in\mc{Q}_j}\left|\mb{h}_{ir}^H\mb{w}_{jr}\right|^2 + \sigma^2} \geq \gamma_i,\forall i \nonumber\\
&&\; \sum_{i\in\mc{K}_q}\left\|\mb{w}_{iq}\right\|^2 \leq \alpha P_q.\nonumber
\end{eqnarray}
In other words, we let all the binary variables $a_{iq}$ to be $1$. Let us denote $\alpha^{\star}$ and $\alpha^{-}$ as
the optimal solutions in problems $\mc{P}_2$ and $\mc{P}_2'$, respectively.

\emph{Remark 4:} Theorem \ref{prop-1} is also applicable to the relaxation of problem $\mc{P}_2$ into problem $\mc{P}_2'$.
Corollary \ref{coro-2} is also applicable to the optimal 
solution of problem $\mc{P}_2'$. Unfortunately, solving problem $\mc{P}_2'$ does not always 
provide us a solution that meets the conditions in Corollary \ref{coro-2}. To illustrate this observation,
let us consider a simple system setting with $K=1$ and $Q=2$. In solving problem $\mc{P}_1'$, the MS 
will be assigned to the BS that requires the lowest transmit power. However, under the problem formulation $\mc{P}_2'$,
the obtained solution will result in \emph{non-zero} transmit powers at both BSs to have $\alpha$ minimized, \emph{i.e.}, the transmit powers
are split and balanced at the both BSs. Nevertheless, in solving problem $\mc{P}_2'$, one can obtain the lower bound on the optimal value of problem $\mc{P}_2$. 

\emph{Remark 5:} Suppose that one has obtained $\{\mb{w}_{iq}^-\}$ as the optimal solution to problem $\mc{P}_2'$. 
Let $q_i^+ = \arg\max_{q\in\mc{Q}_i}\\ \mr{SINR}_{iq} = \arg\max_{q\in\mc{Q}_i} \big|\mb{h}_{iq}^H\mb{w}_{iq}^-\big|^2$ be
the BS association with MS-$i$. Then, for a known BS association profile $\{q_i^+\},i=1,\ldots,K$, an optimal beamforming design 
for minimizing the transmit power margin across the BSs can be easily found \cite{Duy-GC11} by solving the following optimization
\vspace{-0.15cm}
\begin{eqnarray}\label{per-BS-fixed}
&\underset{\{\mb{w}_{iq_i}\},\alpha}{\minimize}&\; \alpha\sum_{q=1}^Q P_q \\
&\st&\; \frac{\big|\mb{h}_{iq_i}^H\mb{w}_{iq_i}\big|^2}{
\sum_{j\neq i}^K\big|\mb{h}_{iq_j}^H\mb{w}_{jq_j}\big|^2 + \sigma^2} \geq \gamma_i,\forall i \nonumber\\
&&\; \sum_{i\in\mc{K}_q,q_i=q}\left\|\mb{w}_{iq_i}\right\|^2 \leq \alpha P_q.\nonumber
\end{eqnarray} 
We denote the obtained per-BS transmit power margin 
as $\alpha^{+}$. 
Certainly, $\alpha^-$ and $\alpha^+$ serve
as a lower bound and an upper bound on $\alpha^{\star}$, \emph{i.e.},
\vspace{-0.15cm}
\begin{equation} \label{bound}
\alpha^- \leq \alpha^{\star} \leq \alpha^+.
\end{equation}
We observe through numerous simulations that the gap between $\alpha^-$ and $\alpha^+$ is nonexistent for most of the simulations (with $K > 1$).
For these cases, solving problem $\mc{P}_2'$ does provide the optimal solution of problem $\mc{P}_2$ too.
For other cases, the BS association profile $\{q_i\},i=1,\ldots,K$ and its corresponding beamforming design
can be employed as a suboptimal solution to problem $\mc{P}_2$. 
\vspace{-0.15cm}
\subsection{QCQP Solution Approach to Problem $\mc{P}_2'$} \label{sect-P-2}
In this section, we apply the QCQP solution approach presented in Section \ref{QCQP-approach} to devise the globally
optimal solution to problem $\mc{P}_2'$. The Lagrangian of problem $\mc{P}_2'$ can be stated as
\vspace{-0.15cm}
\begin{eqnarray}\label{Lagrangian-2}
\hspace{-0.5cm}&&\mc{L}_2(\{\mb{w}_{iq}\},\alpha,\bs{\lambda},\bs{\mu}) \nonumber \\
\hspace{-0.5cm}&&= \alpha\sum_{q=1}^Q P_q 
+ \sum_{q=1}^Q \mu_q\bigg(\sum_{i\in\mc{K}_q}\left\|\mb{w}_{iq}\right\|^2 - \alpha P_q\bigg)\nonumber \\
\hspace{-0.5cm}&&\;\; - \sum_{i=1}^K\!\lambda_i\bigg(\frac{1}{\gamma_{i}}\!\!\sum_{q\in\mc{Q}_i}\!\big|\mb{h}_{iq}^H\mb{w}_{iq}\big|^2 -
\sum\limits_{j\neq i}^K\sum\limits_{r\in\mc{Q}_j}\left|\mb{h}_{ir}^H\mb{w}_{jr}\right|^2 - \sigma^2 \bigg) \nonumber \\
\hspace{-0.5cm}&&= \sum_{i=1}^K\lambda_i\sigma^2 + \alpha\bigg(\sum_{q=1}^Q P_q - \sum_{q=1}^Q \mu_q P_q\bigg) \nonumber \\
\hspace{-0.5cm}&&\;\;+  \sum_{i=1}^K\!\sum_{q\in\mc{Q}_i}\mb{w}_{iq}^H\bigg(\mu_q\mb{I} - \frac{\lambda_i}{\gamma_i}\mb{h}_{iq}\mb{h}_{iq}^H
\!+\! \sum_{j\neq i}^K\!\lambda_j\mb{h}_{jq}\mb{h}_{jq}^H \bigg)\mb{w}_{iq},
\end{eqnarray}
where $\lambda_i$'s and $\mu_q$'s are the Lagrangian multipliers associated with the SINR and the power constraints
and $\bs{\lambda} \triangleq [\lambda_1,\ldots,\lambda_K]^T$ and $\bs{\mu} \triangleq [\mu_1,\ldots,\mu_Q]^T$.
The dual function is then given by $g_2(\bs{\lambda},\bs{\mu}) = \min_{\{\mb{w}_{iq}\},\alpha}\mc{L}_2(\{\mb{w}_{iq}\},\alpha,\bs{\lambda},\bs{\mu})$. 
If any matrix $\mu_q\mb{I} - \frac{\lambda_i}{\gamma_i}\mb{h}_{iq}\mb{h}_{iq}^H
\!+\! \sum_{j\neq i}^K\!\lambda_j\mb{h}_{jq}\mb{h}_{jq}^H$ is not positive semi-definite or 
$\sum_{q=1}^Q P_q < \sum_{q=}^Q\mu_qP_q$, it is possible to find $\mb{w}_{iq}$ or $\alpha >0$ to make 
$g_2(\bs{\lambda},\bs{\mu})$ unbounded below. Thus,
the dual problem is defined as
\vspace{-0.15cm}
\begin{eqnarray} \label{dual-prob-3}
&\underset{\bs{\lambda}\geq \mb{0},\bs{\mu}\geq \mb{0}}{\maximize}&\; \sum_{i=1}^K\lambda_i \sigma^2\\
&\st&\; \mu_q\mb{I} +  \sum_{j\neq i}^K\lambda_j\mb{h}_{jq}\mb{h}_{jq}^H \succeq \frac{\lambda_i}{\gamma_i}\mb{h}_{iq}\mb{h}_{iq}^H,
\forall q \in\mc{Q}_i,\forall i\nonumber \\
&&\; \sum_{q=1}^Q \mu_q P_q \leq \sum_{q=1}^Q P_q. \nonumber
\end{eqnarray}

Compared to problem \eqref{dual-prob-2} with pre-determined weight factors $w_q$'s, the variable $\mu_q$, functioning as the
weight for the transmit power at BS-$q$, have to be optimized in problem \eqref{dual-prob-3}.
Since the dual problem \eqref{dual-prob-3} is convex, its optimal solution can be efficiently obtained 
by standard convex optimization techniques. 

Let $\bs{\lambda}^{\star}$ and $\bs{\mu}^{\star}$ be 
the optimal solution of problem \eqref{dual-prob-3}. 
Except an extremely rare case where the channels from the BSs to the MSs are exactly symmetric,
it is not possible to have $\mu_{q}^{\star}\mb{I} + \sum_{j\neq_i}^K\lambda_j^{\star}\mb{h}_{jq}\mb{h}_{jq}^H \nsucc
\frac{\lambda_i^{\star}}{\gamma_i}\mb{h}_{iq}\mb{h}_{iq}^H,\;\forall q\in\mc{Q}_i,\forall i$.
Thus, $\mc{L}_2(\{\mb{w}_{iq}\},\alpha,\bs{\lambda}^{\star},\bs{\mu}^{\star})$ is a strictly convex function. According
to Property \ref{theorem-3} in Section \ref{QCQP-Sect}, strong duality holds, \emph{i.e.}, the optimal solution of problem $\mc{P}_2'$,
can be found through the dual problem \eqref{dual-prob-3}. It is noted that $\mb{w}_{iq}^-$ must be set to $\mb{0}$
to have the Lagrangian function \eqref{Lagrangian-2} minimized should $\mu_{q}^{\star}\mb{I} + 
\sum_{j\neq_i}^K\lambda_j^{\star}\mb{h}_{jq}\mb{h}_{jq}^H \succ
\frac{\lambda_i^{\star}}{\gamma_i}\mb{h}_{iq}\mb{h}_{iq}^H$. By applying the same argument 
as in the proof of Theorem \ref{theorem-important}, the optimal solution $\{\mb{w}_{iq}^-\}$ 
to problem $\mc{P}_2'$ typically has a sparse structure.

In one special case, among the set of constraints related to a MS, say MS-$i$,
if there exists \emph{only one} $q_i\in\mc{Q}_i$ such that 
\begin{equation} \label{condition-1}
\mu_{q_i}^{\star}\mb{I} + \sum_{j\neq_i}^K\lambda_j^{\star}\mb{h}_{jq_i}\mb{h}_{jq_i}^H \nsucc
\frac{\lambda_i^{\star}}{\gamma_i}\mb{h}_{iq_i}\mb{h}_{iq_i}^H,
\end{equation}
and
\begin{equation} \label{condition-2}
\mu_q^{\star}\mb{I} + \sum_{j\neq}^K\lambda_j^{\star}\mb{h}_{jq}\mb{h}_{jq}^H \succ
\frac{\lambda_i^{\star}}{\gamma_i}\mb{h}_{iq}\mb{h}_{iq}^H,\;\forall q\in\mc{Q}_i, q\neq q_i,
\end{equation}
then one has $\mb{w}_{iq_i}^- \neq \mb{0}$ and $\mb{w}_{iq}^- = \mb{0},\forall q\neq q_i,\forall i$
as the optimal solution to problem $\mc{P}_2'$.
Since the requirements in Corollary \ref{coro-2} are now satisfied, 
the optimal solution of problem $\mc{P}_2$ is also found. In other cases, the steps given in Remark 5 can 
be applied to generate a suboptimal BS association and beamforming design solution of problem $\mc{P}_2$.
We summarize the steps to obtain lower and upper bounds on the optimal per-BS transmit power margin $\alpha^{\star}$
and a suboptimal solution of problem $\mc{P}_2$ in Algorithm \ref{algo-2}.

\begin{algorithm}
\small
Solve the dual problem \eqref{dual-prob-3} to obtain the optimal dual solutions $\bs{\lambda}^{\star}$
and $\bs{\mu}^{\star}$\;
Obtain the optimal primal solution $\{\mb{w}_{iq_i}^-\}$, $\alpha^-$ of problem $\mc{P}_2'$\;
Set $\alpha^-$ as the lower bound on $\alpha^{\star}$\;
Verify if conditions \eqref{condition-1}--\eqref{condition-2} are satisfied\;
\textbf{if} yes\;
\quad \textbf{then} $\{\mb{w}_{iq_i}^-\}$ is optimal to problem $\mc{P}_2$\;
\textbf{otherwise} set $q_i = \arg\max_{q\in\mc{Q}_i}\big|\mb{h}_{iq}^H\mb{w}_{iq_i}^-\big|^2$\;
Set $\{q_i\}$ as a suboptimal BS association strategy\;
Solve problem \eqref{per-BS-fixed} to obtain the upper bound $\alpha^+$ on $\alpha^{\star}$
and the corresponding beamforming design\;
\caption{Iterative Algorithm for Minimizing The Per-BS Transmit Power Margin}
\label{algo-2}
\end{algorithm}
\normalsize
\subsection{A Comparison to the Relaxation-and-Rounding Techniques in \cite{Duy-IET13}} 
In a prior work \cite{Duy-IET13}, we proposed two relaxation-and-rounding techniques to solve the joint BS association
and beamforming design problem $\mc{P}_2$. Of the two techniques, the better performing `relaxation-based-2' approach
first relaxes all BS association variables $\{a_{iq}\}$ to $1$ and finds the beamforming design 
through the optimization
\vspace{-0.1cm}
\begin{eqnarray}\label{opt-prob-5}
\tilde{\mc{P}}_2:\quad &\underset{\{\mb{w}_{iq}\},\alpha}{\minimize}&\; \alpha\sum_{q=1}^Q P_q 
\\
&\st&\; \frac{\bigg|\sum\limits_{q \in \mc{Q}_i}\mb{h}_{iq}^H\mb{w}_{iq}\bigg|^2}{
\sum\limits_{j\neq i}^K\bigg|\sum\limits_{r\in\mc{Q}_j}\mb{h}_{ir}^H\mb{w}_{jr}\bigg|^2 + \sigma^2} \geq \gamma_i,\forall i \nonumber\\
&&\; \sum_{i\in\mc{K}_q}\left\|\mb{w}_{iq}\right\|^2 \leq \alpha P_q.\nonumber
\end{eqnarray}
The obtained solution, denoted as $\{\tilde{\mb{w}}_{iq}\}$, is then utilized for generating a
BS association profile in a similar fashion as given in Remark 5. Subsequently, the beamforming design
can be found accordingly to the generated BS association profile \cite{Duy-IET13}. 

It is noted that there is a subtle difference between problems $\mc{P}_2'$ and $\tilde{\mc{P}}_2$ in 
expressing the SINR constraints. In fact, with the SINR constraint expression in \eqref{opt-prob-5}, 
problem $\tilde{\mc{P}}_2$ mimics the optimization of the beamformers
in a single-cell system with power constraints per groups of antennas.
It is noted that problem $\tilde{\mc{P}}_2$ can be recast into a convex second-order conic program (SOCP) \cite{Duy-IET13}.
A quick verification on the dual problem of the optimization $\tilde{\mc{P}}_2$ would
indicate that the obtained optimal solution $\{\tilde{\mb{w}}_{iq}\}$ does not have a sparse structure, \emph{i.e.},
$\tilde{\mb{w}}_{iq} \neq \mb{0},\,\forall i,\forall q$.
Hence, $\{\tilde{\mb{w}}_{iq}\}$ cannot be an optimal solution of the original problem $\mc{P}_2$, 
unlike the solution $\{\mb{w}_{iq}^-\}$ obtained from solving problem $\mc{P}_2'$. 
In addition, the obtained optimal value from solving problem \eqref{opt-prob-5}, denoted as $\tilde{\alpha}$, is typically much smaller than
the one from solving problem $\mc{P}_2'$, \emph{i.e.}, $\alpha^-$. Thus, the relaxation-and-rounding approach in 
\cite{Duy-IET13} usually generates a large gap between the lower bound and upper bound on $\alpha^{\star}$,
unlike the proposed QCQP solution approach proposed in this work.
A numerical comparison between the two approaches will be presented in the simulation to verify this observation.
\subsection{SDP Relaxation}
In order to obtain a globally optimal solution to the optimization problem $\mc{P}_2'$, 
we can also apply the SDP approach. Let $\mb{X}_{iq} = \mb{w}_{iq}\mb{w}_{iq}^H$ and $\mb{H}_{iq} = \mb{h}_{iq}\mb{h}_{iq}^H$,
the QCQP $\mc{P}_2'$ can be recast as an SDP 
\begin{eqnarray} \label{SDP-2}
\!\!\!\!&\underset{\{\mb{X}_{iq}\},\alpha}{\minimize}& \alpha\sum_{q=1}^QP_q \\
\!\!\!\!&\st& \frac{1}{\gamma_i}\!\!\sum_{q \in \mc{Q}_i}\!\! \tr\{\mb{H}_{iq}\mb{X}_{iq}\} \! - \!
\sum\limits_{j\neq i}^K\!\sum\limits_{r\in\mc{Q}_j}\!\!\tr\{\mb{H}_{ir}\mb{X}_{jr}\!\}\! \geq\! \sigma^2,\forall i \nonumber\\
\!\!\!\!&&\sum_{i\in \mc{K}_q} \tr\{\mb{X}_{iq}\} \leq \alpha P_q \nonumber \\
\!\!\!\!&&\mb{X}_{iq} \succeq \mb{0}. \nonumber 
\end{eqnarray}
Herein, the rank constraint $\mr{rank}\{\mb{X}_{iq}\} = 1$ is dropped to render problem \eqref{SDP-2} as a convex SDP. 
This convex relaxation SDP then can be optimally solved by the interior-point method and a convex SDP solver like \texttt{cvx} \cite{CVX}.

Since strong duality holds for the QCQP $\mc{P}_2'$ and the dual problem of the SDP \eqref{SDP-2} is also 
\eqref{dual-prob-3}, a rank-$1$ solution of $\{\mb{X}_{iq}\},\forall q,\forall i$ can always be found.
Should the obtained solution of problem \eqref{SDP-2} meet all the requirements in Corollary \ref{coro-2}, it is also 
the optimal solution to the  joint optimization of BS association and beamforming design problem $\mc{P}_2$. Otherwise,
the approximation steps in Remark 5 can be applied to generate a suboptimal solution of problem $\mc{P}_2$.
\section{Numerical Results}
This section presents the numerical evaluations on the power consumption
of a multiuser multicell system employing dynamic BS association. In all simulations, 
we assume that the locations of the BSs are fixed and the distance between any two nearby 
BSs is normalized to $1$, whereas the MSs are randomly located. Each BS is equipped with $M=4$ transmit antennas. 
The channel from a BS to a MS is generated from i.i.d. Gaussian random variables (Rayleigh fading) 
using the path loss model with the path loss exponent of $4$ and the reference distance of $1$.
The transmit power at each BS is limited at $1$ W ($0$ dB). The AWGN power spectral density $\sigma^2$ is assumed to be $0.01$ W
while the target SINRs at the MSs are set the same $\gamma$.

\begin{figure}[t!]\centering
    \includegraphics[width=85mm]{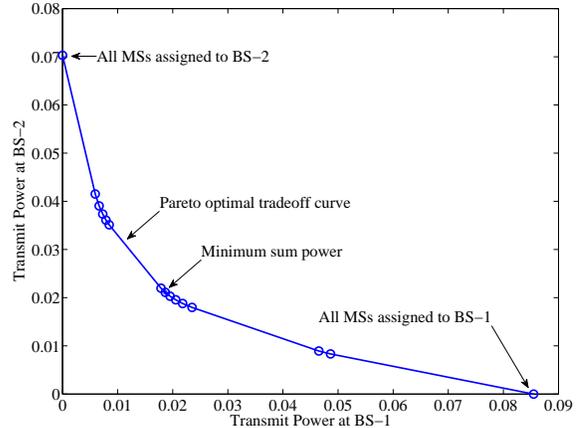}
    \caption{The Pareto-optimal tradeoff curve in power consumption between the two BSs with optimal joint BS association and
    beamforming design.}
    \label{tradeoff}
\end{figure}

In the first simulation setting, we consider a two-cell system with $4$ randomly located MSs 
between the two BSs. The target SINR $\gamma$ is set at $16$ dB. For a randomly generated channel realization, 
we plot in Fig. \ref{tradeoff} the Pareto-optimal tradeoff curve in the transmit powers at the two BSs employing dynamic BS association. 
To obtain each tradeoff point, we vary $w_1$ in the interval $[0,1]$ and set $w_2 = 1-w_1$. 
Depending on the weights, our proposed framework can obtain the corresponding Pareto-optimal joint BS association and beamforming design.
In fact, it is impossible to find a joint BS association and beamforming design that results in a power allocation profile below the plotted 
Pareto-optimal tradeoff curve. Note that at the extreme points of the tradeoff curves, the MSs are all assigned to either one of the two BSs.

\begin{figure}[t!]\centering
	\includegraphics[width=85mm]{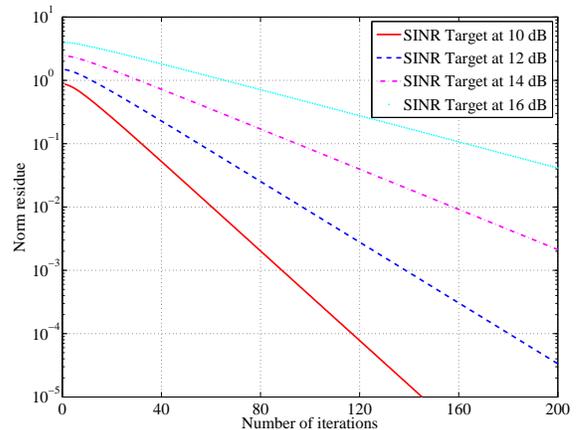}
	\caption{Convergence of the proposed iterative algorithm to solve Problem $\mc{P}_1'$ with different SINR targets. The speed
		of convergence is slower with higher SINR targets.}
	\label{convergence}
\end{figure}

Fig. \ref{convergence} illustrates the convergence of the iterative algorithm in Section \ref{Uplink-Downlink-iteration}, which allows us to 
obtain the optimal solution to problem $\mc{P}_1'$. In the figure, we plot the norm residue $\|\bs{\lambda}^{(n)} - \bs{\lambda}^{\star}\|$ 
(where $\bs{\lambda}^{\star}$ is the optimal uplink power vector) versus the number of iterations with different SINR targets. It is observed that 
the fixed-point iteration \eqref{iteration} converges very fast. Interestingly, the speed of convergence becomes slower with increasing SINR targets.

\begin{figure}[t!]\centering
    \includegraphics[width=60mm]{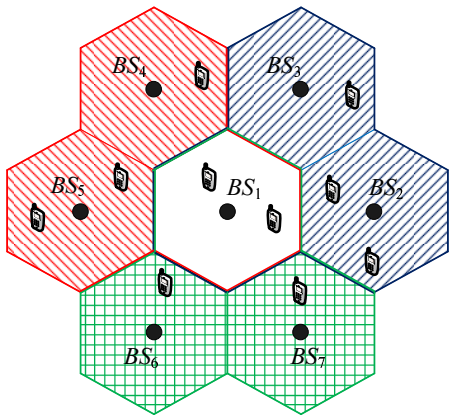}
    \caption{A seven-cell network grouped into three clusters with ten randomly located MSs. A MS can only be associated
    	to one of the BSs in its assigned cluster.}%
    \label{simu-plat}
\end{figure}

In the second simulation setting, we compare the results obtained from the
optimal BS association (with different clustering sizes) to that obtained from 
fixed BS association schemes. Examples of fixed BS association schemes for a MS 
are the channel-based scheme (assigned to the BS with the strongest downlink channel)
and the location-based scheme (assigned to the closest BS). With fixed BS association, 
the beamforming vectors for the MSs and the transmit power at the BSs 
are optimally obtained by means of coordinated beamforming \cite{WeiYu10}. We consider a multicell system with $7$ BSs (each equipped 
with four antennas) and $10$ MSs, as illustrated in Fig. \ref{simu-plat}. Of the 
$7$ cells, we consider two clustering scenarios: i.) universal clustering with all $7$ 
cells and ii.) $3$-cell clustering with cluster \#1 (cell \#1, \#2, and \#3), cluster 
\#2 (cell \#1, \#4, and \#5), and cluster \#3 (cell \#1, \#6, and \#7). In the $3$-cell 
clustering scenario, a MS, say MS-$i$, is first assigned to a cluster based on its 
relative distance to the center of the cluster. MS-$i$ then can only be associated 
to one of the $3$ BSs within its assigned cluster.

\begin{figure}[t!]\centering
    \includegraphics[width=85mm]{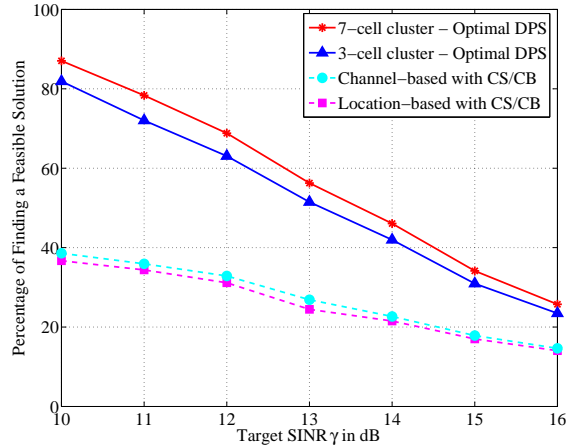}
    \caption{Percentage of finding a beamforming design to meet the target SINR with different BS association schemes. The optimal DPS
    	can double the chance of finding a feasible solution, compared to the fixed BS association schemes with optimal CS/CB \cite{WeiYu10}.}
    \label{percentage}
\end{figure}

\begin{figure}[t!]\centering
	\includegraphics[width=85mm]{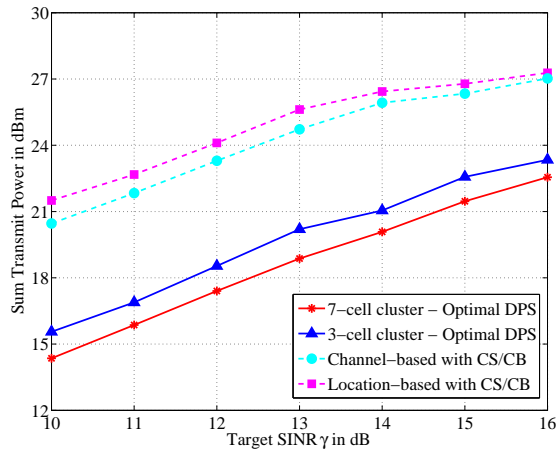}
	\caption{Average sum transmit power \emph{versus} target SINR with different BS association schemes. The optimal DPS
		can save more than $5$ dB in the sum transmit power, compared to the fixed BS association schemes with optimal CS/CB \cite{WeiYu10}.}
	\label{power}
\end{figure}

Fig. \ref{percentage} displays the percentage of finding a feasible beamforming strategy to meet the target SINR
at the MSs with different BS association schemes. As the target SINR varies, $10,000$ channel realizations at each
SINR value are used to obtain the ratios in Fig. \ref{power}. Unlike the first simulation setting with $M=K=4$,
it is not always possible to find a feasible beamforming strategy in the second simulation setting where $M=4$ and $K=10$.
It is observed from the figure that the chance of finding a feasible beamformer design  
can be doubled by the proposed DPS strategy, thanks to 
the optimal and dynamic association of the MSs to the BSs. In contrast, 
by pre-determining the associations, an optimal CS/CB strategy using \cite{WeiYu10}
may not be found at high probability.
Interestingly, by grouping the cells into clusters of $3$ cells, one can obtain nearly the same optimal
performance achieved by the larger cluster of $7$ cells.

\begin{figure}[t!]\centering
	\includegraphics[width=85mm]{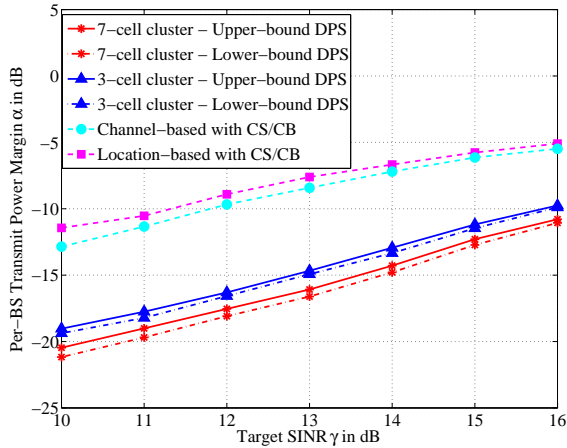}
	\caption{Per-BS transmit power margin \emph{versus} target SINR with different BS association schemes. The optimal DPS
		can provide a much better balance in transmit powers across the BSs and reduce the peak transmit power as much as $7$ dB,
		compared to the fixed BS association schemes with optimal CS/CB \cite{WeiYu10}.}
	\label{margin}
\end{figure}


\begin{figure}[t!]\centering
	\includegraphics[width=85mm]{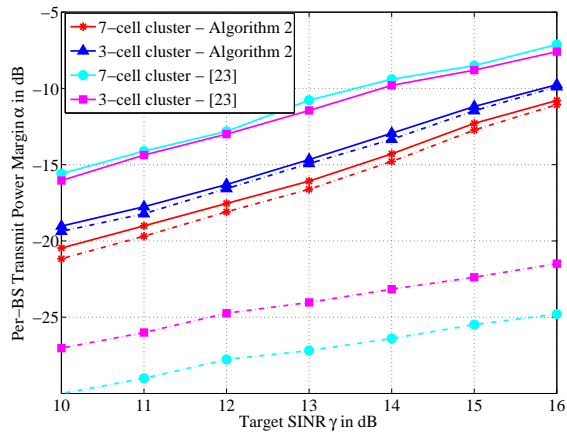}
	\caption{Comparing the per-BS transmit power margins obtained from Algorithm \ref{algo-2} 
		and the approach in \cite{Duy-IET13}: \emph{solid} lines
		are for the upper bounds and \emph{dashed-dotted} lines are for the lower bounds.
		The proposed DPS scheme in Algorithm \ref{algo-2} provides a much smaller gap between its upper bound 
		and lower bound, compared to the approach in \cite{Duy-IET13}.}
	\label{margin-IET}
\end{figure}

Fig. \ref{power} illustrates the average sum transmit power across the $7$ BSs (with equal weights) \emph{versus} the target SINR
at the MSs (each MS is set at the same SINR target). As observed from the figure, more transmit power is required to meet
the higher target SINR. Out of the considered BS association schemes, it is clearly shown 
that the optimal joint BS association and beamforming design significantly outperforms
the fixed BS association schemes (location-based and channel-based). In particular,
the optimal joint schemes can save the transmit power at each BS up to $5$ dB over the fixed BS association schemes
with optimal CS/CB \cite{WeiYu10}.
It is also observed that the optimal joint scheme with 3-cell clustering only imposes a 
penalty of $0.5$ dB in power usage, compared to the full 7-cell clustering. Clearly, a small cluster size is much more beneficial
for practical implementation.


Fig. \ref{margin} shows the performance of the joint BS association and beamforming design for minimizing per-BS transmit power margin. 
As observed from the figure, the per-BS transmit power margin is reduced by at least $5$ dB to $10$ dB by the dynamic BS association schemes
proposed in Algorithm \ref{algo-2}, compared to the fixed BS association schemes with optimal CS/CB in \cite{WeiYu10}. 
Herein, the lower bound was generated by solving problem $\mc{P}_2'$, 
whereas the upper bound was generated by the BS association profile $\{q_i^+\}$ accordingly to the solution of problem $\mc{P}_2'$.
It is also observed from the figure that the gap between the two bounds on the transmit power margin as given in \eqref{bound} 
is very tight for both $3$-cell and $7$-cell clustering schemes. Hence,
the proposed joint BS association and beamforming design in Algorithm \ref{algo-2} 
can generate an exceptionally well-performed and near-optimal solution to the original problem $\mc{P}_2$.


\begin{figure}[t!]\centering
	\includegraphics[width=85mm]{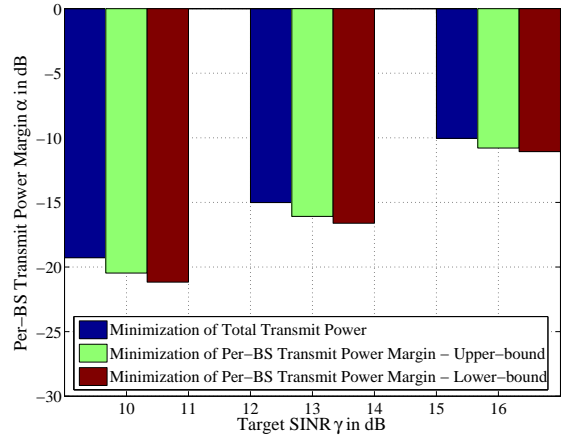}
	\caption{Comparing the per-BS transmit power margin with $3$-cell clustering for the sum power minimization and 
		the per-BS transmit power margin minimization. The latter design criterion can 
		reduce the peak transmit power by $1$-$2$ dB, compared to the former one.}
	\label{compare}
\end{figure}

In Fig. \ref{margin-IET}, we compare the performance between the proposed Algorithm \ref{algo-2} in this work
and the prior work in \cite{Duy-IET13}. As observed from the figure, by relying on the solution of problem
$\tilde{\mc{P}}_2$, the approach in \cite{Duy-IET13} generates a very large gap between the lower bound and 
upper bound on the optimal value of problem $\mc{P}_2$. In contrast,
the tight gap generated by Algorithm \ref{algo-2} allows us to determine 
the minimum per-BS transmit power margin more properly. In addition, coupled with a closer upper bound,
Algorithm \ref{algo-2} also generates a better suboptimal BS association and beamforming design than the approach in \cite{Duy-IET13}.

Finally, Fig. \ref{compare} compares the per-BS transmit power margins with $3$-cell clustering
obtained from the two design objectives: sum power minimization and 
per-BS power margin minimization. It is observed from the figure that the per-BS power
margin can be reduced around $1$-$2$ dB by the latter design criterion.

\section{Conclusion}
This paper has presented a solution framework to obtain an optimal joint BS association
and beamforming design for downlink transmission.
The design objective was to minimize either the weighted transmit power across the
BSs or the per-BS transmit power margin with a set of target SINRs at the MSs.
By properly relaxing the nonconvex joint BS association and beamforming design
problems, we have shown that their optimal solutions
can be obtained via the relaxed problems. Under the first design objective,
such optimality is always guaranteed.
Two solution approaches based on the Lagrangian duality and the dual uplink problem
have been then proposed to find an optimal solution.
Under the second design objective, based on the obtained solution from the relaxed problem,
a near-optimal solution to the original problem is then proposed.
Simulation results have shown the superior performance of the optimal joint BS association and
beamforming design over fixed BS association schemes. In addition, simulation shows that $3$-cell clustering
is sufficient to obtain a very close performance to the universal clustering.


\balance

\balance

\end{document}